\documentclass[12pt]{article}
\usepackage{axodraw} 
\usepackage{amsmath}
\usepackage{verbatim}
\usepackage{latexsym}
\usepackage{graphicx}

\begin{document}

\author{D.~Bardin, L.~Kalinovskaya, V.~Kolesnikov and W. von Schlippe}

\title{$J_{AW,WA}$ functions in Passarino--Veltman reduction}

\date{\today}

\maketitle

\begin{abstract}{
In this paper we continue to study a special class of Passarino--Veltman functions $J$ arising 
at the reduction of infrared divergent box diagrams. We describe a procedure of separation of
two types of singularities, infrared and mass singularities, which are absorbed in simple $C_0$
functions. The infrared divergences of $C_0$'s can be regularized then by any method: 
photon mass, dimensionally or by the width of an unstable particle.
Functions $J$, in turn, are represented as certain linear combinations of the standard 
$D_0$ and $C_0$ Passarino--Veltman functions. The former are free of both types of singularities
and are expressed as explicit and compact linear combinations of logarithms and dilogarithm
functions. We present extensive comparisons of numerical results with those obtained with the aid 
of the LoopTools package.}
\end{abstract}

\newcommand{\bq}{\begin{equation}}
\newcommand{\eq}{\end{equation}}
\newcommand{\bqa}{\begin{eqnarray}}
\newcommand{\eqa}{\end{eqnarray}}
\newcommand{\nll}{\nonumber\\}
\newcommand{\Litwo}{\mbox{${\rm{Li}}_{2}$}}
\newcommand{\ds }{\displaystyle}
\newcommand{\sss}[1]{\scriptscriptstyle{#1}}
\def\Qs{Q^2}
\def\Ts{T^2}
\def\Us{U^2}
\def\Ps{T^2}
\def\P {P^2}
\def\ieps{i\epsilon}
\def\mgm{m_{\gamma}}
\def\mup{m_u}
\def\mdn{m_d}
\def\mbt{m_b}
\def\mtp{m_t}
\def\mw {M_{\sss{W}}}
\def\mws{M_{\sss{W}^2}}
\def\mtpll{m_{t}^2}
\def\mbtll{m_{b}^2}
\def\mupll{m_{u}^2}
\def\mdnll{m_{d}^2}
\def\mwll{M_{\sss W}^2}
\def\Psut{T^2_{13}}

\newcommand{\yPsl}{y_{T_1}}
\newcommand{\yPsll}{y_{T_2}}
\newcommand{\ylO}{y_{l_0}}
\newcommand{\ydO}{y_{d_0}}
\newcommand{\yd}{y_d}
\newcommand{\yl}{y_l}
\newcommand{\mfll}{m_f^2} 
\newcommand{\muno}{m_{1}} 
\newcommand{\mdue}{m_{2}}
\newcommand{\mtre}{m_{3}}
\newcommand{\mqat}{m_{4}}

\newcommand{\Psy}{T^2_y}
\newcommand{\sdPsy}{\sqrt{D_{T}}}
\newcommand{ \bPsy}{b_{T}}
\newcommand{\yPs}{{y_{T}}}

\newcommand{\sDmax}{s_{D}^{\max}}
\newcommand{\sDmin}{s_{D}^{\min}}
\newcommand{\Nomc}{N_{c}}
\newcommand{\sanc}{{\tt SANC~}}

\newcommand{\ykxyl }{y_{k_1}}
\newcommand{\ykxyll}{y_{k_2}}
\newcommand{\bkxyy }{B_{k}}
\newcommand{\sdkxy }{\sqrt{D_{k}}}

\newcommand{\yLmkl }{y_{L^*_1}}
\newcommand{\yLmkll }{y_{L^*_2}}
\newcommand{\bLmk }{B_{L^*_l}}
\newcommand{\sdLmk }{\sqrt{D_{L^*_l}}}
\newcommand{\sdLmkl}{\sqrt{D_1}}
\newcommand{\sdLmkll}{\sqrt{D_2}}

\newcommand{\yPslL}{y_{1}}
\newcommand{\yPsllL}{y_{2}}
\newcommand{\ykxylR }{y_{k1}}
\newcommand{\ykxyllR}{y_{k2}}
\newcommand{\yPslR}{yPslR}
\newcommand{\kxydy}{k_{xy}^2{_|{_{_y}}}}
\newcommand{\zlm}{y_{l-}}
\newcommand{\zlp}{y_{l+}}
\newcommand{\dt}{D_t}
\newcommand{\cp}{c_{+}}
\newcommand{\cm}{c_{-}}

\clearpage

%%%%%%%%%%%%%%%% JAW paper: Introduction %%%%%%%%%%%%%%%%%%%%%%%%%%%%%
% edits by WvS 27-Jul-2009:                                          %
% [[x]] = delete x; [[x > y]] = replace x by y; [[> x]] = insert x;  %
% more edits 27-Oct-2009; 30-Oct-2009; 11-Dec-2009                   %
%%%%%%%%%%%%%%%%%%%%%%%%%%%%%%%%%%%%%%%%%%%%%%%%%%%%%%%%%%%%%%%%%%%%%%
%
\section{Introduction}
%---------------------
In the standard Passarino--Veltman reduction~\cite{Passarino:1978jh}
of 4-point box functions with an internal photon line connecting two external
lines on the mass shell there appears an infrared and mass singular $D_0$ function
(see, for example,~\cite{Bardin:1999ak}).
A typical example of these diagrams arising in the calculation of one-loop EW corrections
to $f\bar{f}\to ZZ(ZA)$ processes was considered in~\cite{Bardin:2009zz}, where a universal
approach to the calculation of such diagrams was proposed.

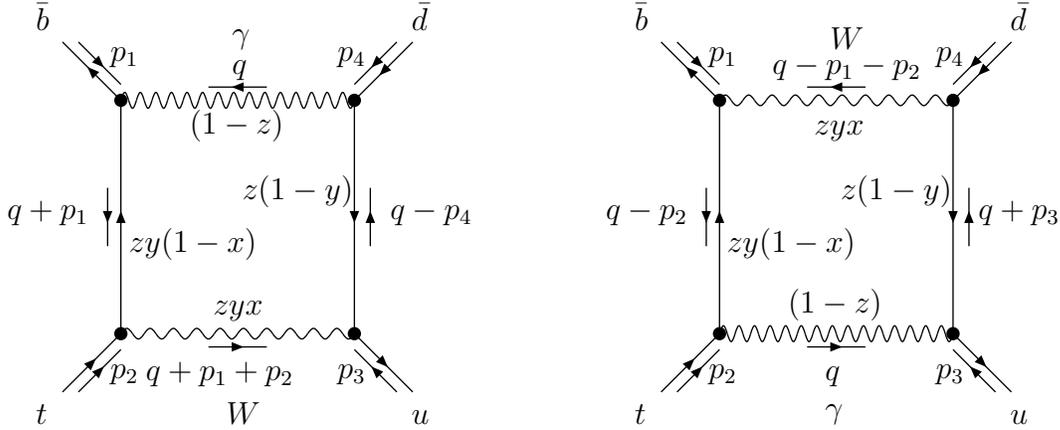
\begin{figure}[!h]
\[
\begin{array}{ccc}
  \vcenter{\hbox{
\begin{picture}(132,132)(-20,0)

\Text(-32,137)[lb]{$\bar b$}
\Text(-32,-12)[lb]{$t$}
\Text(110,137)[lb]{$\bar d$}
\Text(110,-12)[lb]{$u$}

\Text(35,27)[lb]{$zyx   $}
\Text(26,95)[lb]{$(1-z) $}
\Text(46,70)[lb]{$z(1-y)$}
\Text(3,50 )[lb]{$zy(1-x)$}

\Text(40,-12)[lb]{$W$}
\Text(42,118)[lb]{$q$}
\Text(42,130)[lb]{$\gamma$}

\Text(9.5,2)[lb]{$q+p_1+p_2$}
\Text(98,63)[lb]{ $q-p_4$}
\Text(-43,63)[lb]{$q+p_1$}

  \ArrowLine(55,115)(33,115)
  \ArrowLine(33,17)(55,17)

  \ArrowLine(88,110)(88,22)
  \Vertex(88,22){2.5}
  \ArrowLine(88,22)(110,0)
  \Photon(88,110)(0,110){3}{15}
  \Vertex(0,110){2.5}
  \ArrowLine(0,110)(-22,132)
  \ArrowLine(0,22)(0,110)

 \Vertex(0,22){2.5}
 \ArrowLine(-22,0)(0,22)
 \Photon(0,22)(88,22){2}{10}

 \ArrowLine(104,132)(88,116)
 \Vertex(88,110){2.5}
 \ArrowLine(110,132)(88,110)
 \ArrowLine(104,0)(88,16)

 \ArrowLine(-16,0)(0,16)
 \ArrowLine(-16,132)(0,116)

 \Text(-4,123)[lb]{$p_1$}
 \Text(-4,3)[lb]{$p_2$}
 \Text(82,123)[lb]{$p_4$}
 \Text(82,3 )[lb]{$p_3$}

\ArrowLine(94,55)(94,77)

\ArrowLine(-5,77)(-5,55)

\end{picture}}}

& \qquad \qquad \qquad &
  \vcenter{\hbox{
\begin{picture}(132,132)(-20,0)

  \Text(-32,137)[lb]{$\bar b$}
  \Text(-32,-12)[lb]{$t$}
  \Text(110,137)[lb]{$\bar d$}
  \Text(110,-12)[lb]{$u$}

\Text(26,27)[lb]{$ (1-z)  $}
\Text(36,95)[lb]{$ zyx    $}
\Text(46,70)[lb]{$ z(1-y) $}
\Text(3,50 )[lb]{$z y(1-x) $}

\Text(42,130)[lb]{$W$}
\Text(20,118)[lb]{$q-p_1-p_2$}
\Text(40,2)[lb]{$q$}
\Text(40,-12)[lb]{$\gamma$}

  \ArrowLine(55,115)(33,115)
  \ArrowLine(33,17)(55,17)

  \ArrowLine(88,110)(88,22)
  \Vertex(88,22){2.5}
  \ArrowLine(88,22)(110,0)
  \Photon(88,110)(0,110){2}{10}

  \Vertex(0,110){2.5}
  \ArrowLine(0,110)(-22,132)
  \ArrowLine(0,22)(0,110)

 \Vertex(0,22){2.5}
 \ArrowLine(-22,0)(0,22)
 \Photon(0,22)(88,22){3}{15}

 \ArrowLine(104,132)(88,116)
 \Vertex(88,110){2.5}
 \ArrowLine(110,132)(88,110)
 \ArrowLine(104,0)(88,16)

 \ArrowLine(-16,0)(0,16)
 \ArrowLine(-16,132)(0,116)

 \Text(-4,123)[lb]{$p_1$}
 \Text(-4,3)[lb]{$p_2$}
 \Text(82,123)[lb]{$p_4$}
 \Text(82,3 )[lb]{$p_3$}

\Text(98,63)[lb]{$q+p_3$}
\ArrowLine(94,55)(94,77)
\Text(-43,63)[lb]{$q-p_2$}
\ArrowLine(-5,77)(-5,55)
\end{picture}}}
\end{array}
\]
\caption[Remaining $J^d_{AW}$ (left) and $J^d_{WA}$ (right) functions.]
        {Remaining $J^d_{AW}$ (left) and $J^d_{WA}$ (right) functions.}
\label{jawdandjwad}
\end{figure}

In this paper we describe how this approach works for $t\to b f_1\bar{f}'_1$ and
$f_1\bar{f}'_1\to t\bar{b}$ ($f_1$ is a massless fermion) Charged Current (CC) processes.
For these processes one meets eight such box functions, four direct and four cross ones.
Cross boxes are trivially derived from direct ones by a permutation of arguments.
Boxes for $\bar{t}$ decays are related to those of $t$ decays, see~Ref.~\cite{Bardin:2009wv}.
So, it is sufficient to consider only one pair of boxes shown in Fig.~\ref{jawdandjwad}.

\vspace*{5mm}

Using the standard Passarino--Veltman reduction  it is possible to derive relations
between infrared and mass singular functions
$D_0( - \mbtll, - \mtpll, - \mupll, - \mdnll,\Qs,\Ts;0,\mbt,\mw,\mdn)$ and \linebreak
$C_0( - \mdnll, - \mbtll,\Ts;\mdn,0,\mbt)$
and an infrared finite but mass-singular auxiliary function \linebreak
$J^d_{AW}(\Qs,\Ts;\mbt,\mtp,\mdn,\mup,\mw)$ and another
$C_0( - \mupll, - \mdnll,\Qs;\mw,\mdn,0)$ with mass singu\-larity.
These basic relations, exact in masses, are:
\bqa
&&J^d_{AW}(\Qs,\Ts;\mbt,\mtp,\mdn,\mup,\mw)=
\nll
&&\qquad\qquad\qquad\qquad
(\mwll+\Qs) D_0( - \mbtll, - \mtpll, - \mupll, - \mdnll,\Qs,\Ts;0,\mbt,\mw,\mdn)\nll
&&\qquad\qquad\qquad\qquad
 + C_0( - \mupll, - \mdnll,\Qs;\mw,\mdn,0) - C_0( - \mdnll, - \mbtll,\Ts;\mdn,0,\mbt),\qquad
\nll
&&J^d_{WA}(\Qs,\Ts;\mtp,\mbt,\mup,\mdn,\mw)=
\nll
&&\qquad\qquad\qquad\qquad
(\mwll+\Qs) D_0( - \mbtll, - \mtpll, - \mupll, - \mdnll,\Qs,\Ts;\mw,\mtp,0,\mup) \nll
&&\qquad\qquad\qquad\qquad
 + C_0( - \mupll, - \mdnll,\Qs;0,\mup,\mw) - C_0( - \mtpll, - \mupll,\Ts;\mtp,0,\mup).\qquad
\label{basic}
\eqa
Let us emphasize that we have changed the ordering of mass arguments of $J_{WA}^d$ as compared
to $D_0$. For $J_{WA}^d$ they are ordered into two pairs of heavy
($b,t$) and light ($d,u$) quarks such that the first mass in each pair
corresponds to the fermion coupled to the photon, leading to the appearance of a contribution
logarithmically singular in this mass.

The great advantage of basic relations~(\ref{basic}) is the following. The complex object $D_0$,
containing an infrared divergence, is excluded in favor of an explicitly computed $J$ function
and simple objects, $C_0$ functions, whose infrared divergences can be regularized by any 
method: by a photon mass, by dimensional regularization or by the width of an unstable particle.
Examples of the latter $C_0$ functions, regularized by the width, may be found in 
Ref.~\cite{Bardin:2009wv}.

We use the standard \sanc definitions:
\bq
Q^2=(p_1+p_2)^2,\qquad T^2=(p_2+p_3)^2,\qquad U^2=(p_2+p_4)^2.
\eq

The paper is organized as follows.

Section 2 is devoted to the calculation of one of these functions --- $J^d_{AW}\equiv J$.
From here on we omit indices of $J$ since the list of arguments uniquely defines its type.

In Section 3 we present a similar calculation of the $J$ function for the process $ud\rightarrow tb$.
The direct ${AW}$ and ${WA}$ functions are defined by:
\bqa
&&J_{AW}^d(\Qs,\Ps;\mdn,\mup,\mbt,\mtp)=
\nll &&\qquad\qquad\qquad\qquad
(\Qs+\mwll) D_0(-\mdnll,-\mupll,-\mtpll,-\mbtll,\Qs,\Ps; 0,\mdn,\mw,\mbt)
\nll &&\qquad\qquad\qquad\qquad
       +C_0(-\mtpll,-\mbtll,\Qs;\mw,\mbt,0)-C_0(-\mbtll,-\mdnll,\Ps;\mbt,0,\mdn),\qquad
\nll
&&J_{WA}^d(\Qs,\Ps;\mup,\mdn,\mtp,\mbt)=
\nll &&\qquad\qquad\qquad\qquad
(\Qs+\mwll) D_0(-\mdnll,-\mupll,-\mtpll,-\mbtll,\Qs,\Ps;\mw,\mup,0,\mtp)
\nll &&\qquad\qquad\qquad\qquad
       +C_0(-\mtpll,-\mbtll,\Qs;0,\mtp,\mw)-C_0(-\mupll,-\mtpll,\Ps;\mup,0,\mtp).\qquad
\eqa
Again, we limit ourselves to the presentation of function $J_{AW}^d$.

In Section 4 we briefly discuss the $J$ functions for the $t$ channel process $bu\rightarrow td$.

For all processes we take the limit of vanishing light quark masses.
The mass of the quark which is not coupled to the photon may be set equal to zero,
while that for the quark coupled with the photon develops a mass singular logarithm.
We keep logarithmic terms and neglect quark masses everywhere else.
This approximation results in different expressions for $J$ functions for the three channels under
consideration, and this is why the derivation must be presented for three channels separately.

Every Section ends by a numerical comparison with results obtained with the aid of the LoopTools
package~\cite{Hahn:1998yk} for zero width and IR regularization by infinitesimal photon mass.

Section 5 contains a short introduction to the FORTRAN packages, which realize
the calculation of ``doubly subtracted'' $J$ functions (see Sections 2--4 for their definition).

In Section 6 we present our conclusions.

%%%%%%%%%%%%%%%% JAW paper: CalculationJAW_2 %%%%%%%%%%%%%%%%%%%%%%%%%%%
% edits by WvS 27-Jul-2009:                                            %
% [[x]] = delete x; [[x > y]] = replace x by y; [[> x]] = insert x;    %
% [[x ??]] = I do not understand x, please explain in Russian          %
% [[Q(WvS): ... ?]] = a general query by WvS                           %
% some spelling corrections not marked.                                %
%%%%%%%%%%%%%%%%%%%%%%%%%%%%%%%%%%%%%%%%%%%%%%%%%%%%%%%%%%%%%%%%%%%%%%%%
% edits by WvS 23-Oct-2009:                                            %
% (i) corrected typing errors; no [[...]] marks left                   %
%     please note: Feynman has only 1 n at the end!!                   %
% 27-Oct-2009: transferred edits from h/c to file; not all [[]] shown  %
% 30-Oct-2009: definition of Li2; mention FORTRAN;                     %
%%%%%%%%%%%%%%%%%%%%%%%%%%%%%%%%%%%%%%%%%%%%%%%%%%%%%%%%%%%%%%%%%%%%%%%%

\section{Calculation of the $J$ function for $t\to bud$ decay\label{decay}}
%--------------------------------------------------------------------------
\subsection{Representation in the form of a triple integral}
%-----------------------------------------------------------

 The basic definition of the function $J$ reads:
\bq
i\pi^2J(Q^2,T^2;\mbt,\mtp,\mdn,\mup,\mw)=\mu^{4-n}\int d^nq\frac{-2q\cdot p_2}{d_0d_1d_2d_3}\,,
\eq
where
\bqa
d_0 &=& q^2\,,\quad d_1\;=\;(q+p_1)^2+\mbt^2\,,
\nll
d_2 &=& (q+p_1+p_2)^2+\mwll\,,\quad d_3\;=\;(q-p_4)^2+\mdn^2\,.
\eqa

 By standard Feynman parametrization introducing variables $x,y,z$, as 
is shown in Fig.~\ref{jawdandjwad}, one can pass to a $(3+n)$-tuple integral
over  $x,\,y,\,z$ and over the internal momentum $q$.
In $n$-dimensional space we have:
\bqa
   \int\limits_{ }^{ }d^nq \frac{q_\mu} {(q^2 - 2q p + m^2)^\alpha} =
   i\pi^{\frac{\sss n}{\sss 2}}\frac{\Gamma(\alpha - \frac{n}{2})}{\Gamma(\alpha)}
             \bigl(m^2 - p^2)^{\frac{\sss n}{\sss 2} - \alpha} p_\mu\,.
\label{dimreg}
\eqa
 In our case we have $m^2 = Lz$, $p = z k_{xy}$, $\alpha = n = 4$,
 therefore ~Eq.(\ref{dimreg}) becomes:
\bqa
 \int\limits_{ }^{ }d^4 q \frac{q_{\mu}\;(p_4)_\mu} {\bigl(q^2 - 2zq k_{xy} + Lz\bigr)^4} =
    i\pi^2\frac{\Gamma(2)}{\Gamma(4)}
                        \bigl(Lz - z^2 k_{xy}^2 \bigr)^{-2} z (k_{xy})_{\mu}\;(p_4)_\mu\,.
\eqa
%---
In terms of Feynman variables the denominator takes the form:
\bqa
D= d_0 (1-z) + d_1 zy(1-x) + d_2 zxy + d_3 z (1-y).
\label{Deno}
\eqa
From expression~(\ref{Deno}) we derive:
\begin{equation}
   D = q^2 - 2zq k_{xy} + Lz - i\epsilon,
\end{equation}
where the variable $L$ and the vector $k_{xy}$ are:
\bqa
L    &=& (\Qs+\mw^2) x y+\left(p_1^2+\mbt^2 \right) (1-x) y+\left( p_4^2+\mdn^2 \right) (1-y),
\nll
k_{xy} &=& p_4(1-y)-p_1 y(1-x)-(p_1+p_2) x y.
\eqa

Since $-i\epsilon$ is an infinitesimal addition, it is possible to replace it
by $-i\epsilon z$ and redefine $L$ and $D$:

\bqa
   L = L - i\epsilon,
\nll
   D = q^2 - 2zq k_{xy} + Lz.
\eqa
The triple integral over Feynman parameters may be expressed by the same
Eqs.~(13),~(16)--(17) as given in  Ref.~\cite{Bardin:2009zz}:
\bqa
J = \int \limits_{0}^{1}dx\int\limits_{0}^{1}y\,dy
  N_{xy} \int\limits_{0}^1 dz \frac{z}{(L - z k_{xy}^2)^2}\,,
\label{JAWint3}
\eqa
where we have neglected the light quark mass $\mup$ which does
not lead to a mass singularity, and changed the notation of
the other masses as follows:
\bqa
\mbt & \to & \muno\,,
\nll
\mtp & \to & \mdue\,,
\nll
\mw  & \to & \mtre\,,
\nll
\mdn & \to & \mqat\,.
\label{mass_redenote}
\eqa
The ingredients entering Eq.~(\ref{JAWint3}) are
\bqa
N_{xy}&=&  -2 k_{xy} p_2 = -2\mdue^2 x y  +  N_y\,,
\nll
N_{y} &=&  \Ts(1-y) + y\mdue^2 + \Qs + \muno^2\,,
\eqa
and
\bqa
k_{xy}^2&=& -\mdue^2 x^2y^2 + N_y x y - T^2_y\,,
\nll
L       &=& \P xy - i\epsilon,
\nll
\P      &=& \Qs + \mtre^2\,,
\nll
%\eqa
%and the quadratic trinomial $ T^2_y$ is:
%\bqa
T^2_y  &=& \Ts y(1-y) + \muno^2 y + \mqat^2 (1-y).
\eqa
%---

\subsection{Integration with respect to $z$}
%-------------------------------------------
The integration with respect to $z$ is straightforward:
\bqa
&& \int\limits_{0}^{1}dz  \frac{z }{(L - z k_{xy}^2)^2}
=   \frac{1}{(k_{xy}^2)^2}
   \left[\ln(L - k_{xy}^2) - \ln L + \frac{k_{xy}^2}{L - k_{xy}^2} \right].
\label{J_z_integration}
\eqa

\subsection{Integration with respect to $x$}
%--------------------------------
In Eqs.~(\ref{JAWint3}),~(\ref{J_z_integration}) we perform, first of all,
 a change of variables:
\bqa
xy=x^{'}, \quad y dx=dx^{'},
\eqa
hence the ingredients become:
\bqa
N_{xy}   &=& -2 \mdue^2 x  +  N_y\,,
\nll
k_{xy}^2 &=& -\mdue^2 x^2 + N_y x - T^2_y\,,
\nll
L &=& \P x - \ieps\,.
\eqa

{The key identity} is
\bqa
 N_{xy}&=&\frac{d}{dx} k_{xy}^2\,;
\eqa
it allows the integration by parts in full analogy with section 2.3 of Ref.~\cite{Bardin:2009zz}:
\bq
  I(y) = \int\limits_{0}^{y} dU V = UV \Biggl|_\delta^y - \int\limits_{\delta}^y dV U,
\label{intpart}
\eq
where we have introduced an infinitesimal parameter $\delta$,
because both parts in Eq.~(\ref{intpart}) diverge separately.

Let
\bq
L^* = L - k_{xy}^2\,.
\eq
be a quadratic trinomial in $x$:
\bq
L^* = A_{L}x^2 + B_{L}x + C_{L} = A_{L}(x - x_{L_1})(x - x_{L_2}),
\eq
with coefficients
\bqa
A_{L}   &=& \mdue^2\,,
\nll
B_{L}   &=&  (\Ps-\mdue^2) y -\Ps-\muno^2+\mtre^2\,,
\nll
C_{L}   &=&  \Ps y (1-y) + \muno^2 y + \mqat^2 (1-y)-\ieps\,,
\label{QT}
\eqa
and discriminant
\bqa
D_{L}   &=& B_{L}^2-4 A_{L} C_{L}\,.
\label{QTdisc}
\eqa
%---

Next, we introduce the following notation:
$L^*{_|{_{_y}}}=L^*(x = y,\,y)$, $L^*{_|{_{_0}}}=L^*(x = 0,\,y)$ and
$k_{xy}^2{_|{_{_y}}}=k_{xy}^2(x = y,\,y)$.
The two binomials are
\bqa
L^*{_|{_{_y}}} &=& \mtre^2 y+\mqat^2 (1-y)-\ieps\,,
\nll
{k_{xy}^2{_|{_{_y}}}} &=& \Qs y-\mqat^2 (1-y)\,.
\label{binomials}
\eqa
For Eq.~(\ref{intpart}) one has
\bq
dU=\frac{dk_{xy}^2}{(k_{xy}^2)^2}\,,\qquad
\frac{dV}{dx}=\frac{\P k_{xy}^2}{LL^*}-\frac{k_{xy}^2}{(L^*)^2}\frac{dL^*}{dx}\,,
\eq
and the integral $I(y)$ becomes:
\bq
I(y)=-\frac{1}{k_{xy}^2{_|{_{_y}}}}\left[\ln(L^*{_|{_{_y}}})-\ln(L{_|{_{_y}}})\right]
     -\frac{1}{T^2_y}\left[\ln(L^*{_|{_{_0}}})-\ln(L{_|{_{_\delta}}})\right]
+\int^y_{\delta}\frac{\P dx}{LL^*}\,.
\label{intx}
\eq
After some more calculations we arrive at a one-dimensional integral over $y$,
where the infinitesimal parameter $\delta$ cancels out.

%\clearpage

\subsection{Integration over $y$}
%--------------------------------
We proceed with the one-dimensional integral:
\begin{equation}
   J = \int\limits_{0}^{1}\,dy I(y),
\end{equation}
with integrand
\bqa
I(y) = -\frac{1}{k_{xy}^2{_|{_{_y}}}}\Bigl[\ln(L^*{_|{_{_y}}}) - \ln(\P y)\Bigr]
   +\frac{1}{2\sqrt{D_{T}}}\left(\frac{1}{y-\yPsl} - \frac{1}{y-\yPsll}\right)I_{p}(y)\,,
\label{IntegrOvery}
\eqa
and
\bqa
I_{p}(y) &=&
        \ln(C_{L})+\ln(L^*{_|{_{_y}}})-2\ln(\P y) \nll
       && + \frac{B_{L}}{\sqrt{D_{L}}}\left[
               \ln\left(\frac{2 C_{L}+y\left(B_{L}+\sqrt{D_{L}}\right)}{C_{L}}\right)
              -\ln\left(\frac{2 C_{L}+y\left(B_{L}-\sqrt{D_{L}}\right)}{C_{L}}\right)
                                      \right].
\label{Ipofy}
\eqa
Here $P$ gets redefined:
\bq
\P = \Qs+\mtre^2-\ieps\,.
\label{Predefined}
\eq

The quadratic trinomial $C_L$ (see~(\ref{QT})),
\bqa
C_{L}&=& a_{T} y^2+b_{T} y +c_{T},
\nll
a_{T}&=&-\Ts,
\nll
b_{T}&=& \Ts + \muno^2 - \mqat^2\,,
\nll
c_{T}&=& \mqat^2 - \ieps\,,
\nll
D_{T}&=& b^2_{T} - 4 a_{T}c_{T}\,,
\label{yTroots}
\eqa
%---
has the roots:
\bqa
y_{T_{1,2}} &=&\frac{b_{T}\pm\sdPsy}{2\Ts}\,.
\eqa
The following steps of integration with respect to $y$ deviate from the presentation given in Ref.~\cite{Bardin:2009zz}.

\subsubsection{Splitting into three parts}
%-----------------------------------------

Let us redistribute terms in Eqs.~(\ref{IntegrOvery})--(\ref{Ipofy}) into three parts:
\bq
I(y)=I_0(y)+I_1(y)+I_2(y),
\eq
with
\bqa
I_0(y) &=&
\left[-\frac{1}{\kxydy}+\frac{1}{\sdPsy}\left(\frac{1}{y - y_{T_1}}-\frac{1}{y - y_{T_2}}\right)\right]
\left[ \ln\left(1-\frac{y}{y_{l0}}\right)-\ln\left(\frac{\P y-\ieps}{\mqat^2-\ieps}\right)\right],
\nll
I_1(y_1)&=&\frac{1}{2\sdPsy}\frac{1}{y_1}
\bigg[ N(y_1) +\frac{C_1}{\sqrt{D_1}} \left( M(y_1,\sqrt{D_1}) - M(y_1,-\sqrt{D_1}) \right) \bigg],
\nll[1mm]
I_2(y_2)&=&-I_1(y_1).
\label{allContributions}
\eqa
For $I_1(y_1)$ and $I_2(y_2)$ we have changed the variables,
\bq
y_{i} = y - y_{T_i}\,,
\eq
and used the notation
\bqa
D_i &=&D_{L| y \rightarrow y_i}=A_i^2 y_i^2-2 B_i y_i+C_i^2\,,
\nll
A_i &\equiv& A  = \Ps+\mdue^2\,,
\nll
B_i &=& - A^2 \yPs_i+A (\Ps+\Delta_{13})+2 \mdue^2 \Delta_{34}\,,
\nll
C_i &=& (\Ps-\mdue^2) \yPs_i-\Ps-\Delta_{13}\,.
\eqa
Furthermore,
\bqa
C(y_1) &=& -y_1\left(\Ps y_1+\sdPsy\right)=C(y_2)=-y_2\left(\Ps y_2-\sdPsy\right),
\nll
N(y_i) &=& \ln\left(\frac{C(y_i)}{\mqat^2-\ieps}\right)-\ln\left(1-\frac{y}{y_{l0}}\right),
\nll
M(y_i,\sqrt{D_i}) &=& \ln\left(
\frac{C(y_i)-\mdue^2 (y_i+y_{T_i})^2+\mqat^2-\ieps+(y_i+y_{T_i}) (\Delta_{34}+\sqrt{D_i})}
     {C(y_i)}        \right),
\label{ABCDi}
\eqa
and
\bqa
y_{l0} &=& -\frac{(\mqat^2-\ieps)}{\Delta_{34}}\,, \nll
y_d    &=& \frac{\mqat^2}{\Qs+\mqat^2}\,,          \nll
\kxydy &=& (\Qs+\mqat^2) (y-y_d)\,,                \nll
\Delta_{ij}&=& m_i^2-m_j^2\,.
\eqa
These expression are still valid for all $J's$ and are exact in $\mqat$.
The main idea of this splitting arises from the observation
that the $\mqat$ singularities are completely confined to $I_0$,
hence when calculating $I_1$ and $I_2$
one may take the limit $\mqat\to 0$ before taking the integrals.

\subsubsection{Part $J_0$}
%-------------------------
The integral for $J_0$ is straightforward, and we limit ourselves to
presenting the answer in the limit $\mqat\to 0$:
\bqa
J_0&=&\int^1_0I_0(y)dy=
\frac{1}{\Qs}\biggl[\ln\left(\frac{\P}{\mtre^2}\right)\ln\left(-\frac{\Qs}{\mqat^2}\right)
  +\Litwo\left(\frac{\P}{\mtre^2}\right)-\zeta(2)\biggr]
\nll &&
     +\frac{1}{\sdPsy}\biggl\{
      -\ln\left(\frac{\P}{\mtre^2}\right)\ln\left(\frac{\mtre^2}{\mqat^2}\right)
      -\ln\left(\frac{\P}{\mtre^2}\right)\ln\left(R_{13}\right)
\nll &&
      +\ln\left(\frac{\P}{\mtre^2}\right)\ln\left(1-\frac{\Ps}{\Ps+\muno^2-\ieps}\right)
      -\Litwo\left(1-R_{13}\right)
      +\zeta(2)\biggl\}\,,
\eqa
where
\bq
R_{1i} = \frac{\ds \Ps+\muno^2-\ieps}{\ds m_i^2}\,,\quad i=1,3\,
\label{R1i}
\eq
and $\Litwo(x)$ is the dilogarithm function defined by 
${\displaystyle \Litwo(x)=-\int_{0}^{1} \frac{\ln|1-xt|}{t}\,dt}$.

\subsubsection{Part $J_1$}
%-------------------------
For $J_1$ we directly take the limit $\mqat=0$ in which it becomes
\bq
J_1=\int^1_0I_1(y)dy =\int^1_0dy\frac{1}{2\sdPsy}\frac{1}{y}
\bigg[\ln\left(\frac{\Ps (1-y)+\muno^2-\ieps}{\mtre^2-\ieps}\right)
+\frac{C_1}{\sqrt{D_1}}\ln\left(R(y)\right)\bigg],
\label{I_1}
\eq
with the ratio
\bq
R(y)=
\frac{\Psut- A_1 y +\sqrt{D_1}}
     {\Psut- A_1 y -\sqrt{D_1}}\,,
\label{ratio_1}
\eq
where, since we neglect $\mqat$,
\bqa
D_1&\equiv& D_1(y) = A_1^2 y^2 - 2 B_1 y + C_1^2\,,
\nll
A_1&\equiv& A  = \Ps+\mdue^2,
\nll
B_1&=& A\Psut-2\Ps(\mtre^2-\ieps),
\nll
%C_1^2&=&(\Ps+\muno^2-\mtre^2)^2+4\mdue^2\ieps\,.
C_1&=&-(\Ps+\muno^2-\mtre^2)\,,
\nll
\Psut&=&\Ps+\muno^2+\mtre^2-2\ieps\,.
\label{ingredients_1}
\eqa

%One can prove that such manipulations with logarithmic functions
%are allowed by signs of the imaginary parts of their arguments.

Then we use the substitution:
\bq
\sqrt{D_1(y)}=C_1+\frac{Ex-B_1}{C_1}y\,,
\label{subst_JAW1}
\eq
where
\bq
E=C_1\sqrt{D_1(y)_{|y=1}}-C_1^2+B_1\,.
\eq
From Eq.~(\ref{subst_JAW1}) we have:
\bqa
y&=&2\frac{C_1^2Ex}{d(x)}\,,
\nll
\sqrt{D_1(y)}&=&\frac{C_1 n(x)}{d(x)}\,,
\nll
\frac{dy}{dx}&=&2\frac{C_1^2En(x)}{d^2(x)}\,,
\nll
\frac{dy}{y\sqrt{D_1(y)}}&=&\frac{dx}{C_1 x}\,,
\eqa
where
\bqa
d(x)&=&A^2C_1^2-(Ex-B_1)^2\,,
\nll
n(x)&=&A^2C_1^2-B_1^2+E^2x^2\,.
\eqa

In order to take the integral~(\ref{I_1}) we need to know
%\bq
%\frac{dy}{y}=\frac{n(x)dx}{d(x)x}=dx\left(\frac{1}{x}+\frac{E}{AC_1+B_1-Ex}-\frac{E}{AC_1-B_1+Ex}\right),
%\eq
the limits of variation of $x$. Let
\bq
\delta \leq y \leq 1\,,
\eq
then
\bq
x_{|_{\delta}}\equiv\frac{A^2C_1^2-B_1^2}{2C_1^2E}\delta \leq x \leq 1\,.
\eq
Next, replacing the variable $y$ in the ratio~(\ref{ratio_1}) by $x$, we find
%\bq
%R(x) = \frac{((B_1+AC_1)-Ex)((\Psut+C_1)(B_1-AC_1)-E(\Psut-C_1)x)}
%            {((B_1-AC_1)-Ex)((\Psut-C_1)(B_1+AC_1)-E(\Psut+C_1)x)}\,,
%\eq
%or, equivalently,
\bq
R(x) = \frac{(\Psut+C_1)[1-Ex/(B_1+AC_1)][1-Ex(\Psut-C_1)/(\Psut+C_1)/(B_1-AC_1)]}
            {(\Psut-C_1)[1-Ex/(B_1-AC_1)][1-Ex(\Psut+C_1)/(\Psut-C_1)/(B_1+AC_1)]}\,.
\eq
%--

 Taking account of all above equations, we rewrite the integral $J_1$ of Eq.~(\ref{I_1})
using both variables $x$ or $y$ where convenient:
\bqa
J_1 &=&\frac{1}{2\sdPsy}\Biggl\{
 \ln\left(\frac{\Ps+\muno^2-\ieps}{\mtre^2-\ieps}\right)\int^1_{\delta}\frac{dy}{y}
+\int_0^1\frac{dy}{y}\ln\left(1-\frac{\Ps y}{\Ps+\muno^2-\ieps}\right)
\nll& &
+\ln\left(\frac{\Psut+C_1}{\Psut-C_1}\right)\int^1_{x_{|_{\delta}}} \frac{dx}{x}
\nll& &
+\frac{\Psut}{C_1}\int^1_{0}\frac{dx}{x}\biggl[
-\ln\left(1-\frac{Ex}{(B_1+AC_1)}\right)
-\ln\left(1-\frac{E(\Psut-C_1)x}{(\Psut+C_1)(B_1-AC_1)}\right)
\nll& &\hspace{24.5mm}
+\ln\left(1-\frac{Ex}{(B_1-AC_1)}\right)
+\ln\left(1-\frac{E(\Psut+C_1)x}{(\Psut-C_1)(B_1+AC_1)}\right)\biggr]\Biggr\}\,.\quad
\label{I_1_xy}
\eqa
%---
Finally we get
\bqa
 J_1&=&\frac{1}{2\sdPsy}\Biggl\{
 \ln\left(\frac{\Ps+\muno^2}{\mtre^2}\right)\ln\left(\frac{A^2C_1^2-B_1^2}{2C_1^2E}\right)
\nll &&
    -\Litwo\left(\frac{\Ps}{\Ps+\muno^2}\right)
    -\Litwo\left(\frac{E}{B_1+AC_1}\right)
    -\Litwo\left(\frac{E(\Psut-C_1)}{(\Psut+C_1)(B_1-AC_1)}\right)
\nll &&
    +\Litwo\left(\frac{E}{B_1-AC_1}\right)
    +\Litwo\left(\frac{E(\Psut+C_1)}{(\Psut-C_1)(B_1+AC_1)}\right)
\Biggr\}\,.
\label{J1ala2006}
\eqa

\subsubsection{Part $J_2$}
%-------------------------
For $I_2(y_2)$ of (\ref{allContributions}) we set
\bq
y_2=y-\yPs_2\,,
\eq
where in the limit $\mqat\to 0$
\bq
\yPs_2=1+\frac{\muno^2-\ieps}{\Ps}\,.
\eq

$\bullet$ {Transition from variable $y_{2}$ to variable ${t_{2}}$}\\

At this step we make the transition from the variable $y_{2}$ to ${t_{2}}$ by
\bq
\sqrt{D_2}=\sqrt{A^2 y_2^2-2 B_2 y_2+C_2^2}=|C_2|+y_2 t_2\,.
\label{td}
\eq

\noindent
The presence of $|C_2|$ leads to the appearance of two branches,
$|C_2|=\pm C_2 > 0$, in the final answer for $J_2$.
For our definitions see Eqs.~(\ref{ABCDi}).

As a consequence of (\ref{td}) we get
\bqa
y_2 = 2 \frac{B_2+C_2t_2}{A_2^2-t_2^2}\,, \qquad \sqrt{D_2}=\frac{N_2}{A_2^2-t_2^2}\,,
\eqa
with
\bqa
N_2=A_2^2 C_2 + 2 B_2 t_2+C_2t_2^2\,.
\eqa
The Jacobian of the transition is
\bqa
\frac{dy_2}{dt_2}=2\frac{N_2}{(A^2-t_2^2)^2}\,.
\eqa
We also need
\bq
\frac{dy_2}{y_2}=\frac{N_2dt_2}{(B_2+C_2t_2)(A^2-t_2^2)}
                =\left(\frac{C_2}{B_2+C_2t_2}+\frac{1}{A-t_2}-\frac{1}{A+t_2}\right)dt_2\,,
\eq
and
\bq
\frac{dy_2}{y_2\sqrt{D_2}}=\frac{dt_2}{B_2+C_2t_2}\,.
\eq
The limits of integration are
%metka091027
\bqa
t_2^{\max} &=& \frac{\sDmax-C^\pm_2}{1-y_{T2}},\qquad
t_2^{\min}\;=\;\frac{\sDmin-C^\pm_2}{ -y_{T2}},\nll
\eqa
The remaining variables read
\bqa
\sDmax &=& \sqrt{(\mdue^2+\Delta_{13})^2-4 \mdue^2 (\muno^2-\ieps)}\,,
\nll
\sDmin &=& \sqrt{(\Ps    +\Delta_{13})^2-4 \mdue^2 (\mqat^2-\ieps)}\,.
\eqa

$\bullet$ {Replacement of variable from $t_2$ to $y'\equiv y$.}\\

The next replacement reads:
\bqa
t_2=D_{t2}y +t_{2}^{\min},
\eqa
where
\bq
D_{t2} = t_{2}^{\max}-t_{2}^{\min}.
\eq
The Jacobian of this transition is
\bq
\frac{dt_2}{B_2+C_2 t_2}=\frac{dy}{C_2(y-y_{d2})}\,.
\eq

The root $y_{d2}$ is labelled by a second index $\pm$,
depending on the sign of $C^\pm_2$.
\bqa
y^\pm_{d2}   = - \frac{B_2+C^\pm_2 t_2^{min}}{C^\pm_2 D_{t2}}.
\eqa

For $I_2(y_2)$ we get from Eq.~(\ref{allContributions})
\bqa
I_2(y_2)=
-\frac{1}{2\sqrt{D_T}y_2}\left\{
N(y_2)+\frac{C_2^\mp}{\sqrt{D_2}}\left[ M(y_2,\sqrt{D_2})-M(y_2,-\sqrt{D_2})\right]
\right\},
\label{I_2}
\eqa
and
\bq
C_2^\mp = \pm C_2.
\eq

The second term in curly brackets of (\ref{I_2}), after replacement of variables~(\ref{td}), reads:
\bqa
\frac{C_2^+}{y_2 \sqrt{D_2}}
    \left[ M(y_2,\sqrt{D_2})-M(y_2, -\sqrt{D_2})  \right] &=&
  \frac{C_2^+}{y_2\sqrt{D_2}}
        \left[  \ln\left( \frac{n_2^{++}}{d_2^+}\right)
               -\ln\left( \frac{n_2^{+-}}{d_2^+}\right) \right],
\nll
\frac{C_2^-}{y_2 \sqrt{D_2}}
    \left[ M(y_2, \sqrt{D_2})-M(y_2, -\sqrt{D_2})  \right] &=&
  \frac{C_2^-}{y_2\sqrt{D_2}}
        \left[  \ln\left( \frac{n_2^{-+}}{d_2^-}\right)
               -\ln\left( \frac{n_2^{--}}{d_2^-}\right) \right],
\eqa
%---
\bqa
d_2^{\pm}  &=& - 2\Ps (B_2+C_2^\pm t_2),
\nll
n_2^{-+/+-}&=& \pm 2\left(B_2+C_2^\mp t_2\right)\left(t_2\mp A\right),
\nll
n_2^{++/--}&=& \pm 2\left(B_2\mp AC_2^\pm\right)\left(t_2\mp A\right).
\eqa

$\bullet$ {Solution $C_2^-$}
%---------------------------
\bq
I_2(y_2)=
-\frac{1}{2\sqrt{D_{T}}y_2}\left\{
N(y_2)+\frac{C_2^-}{\sqrt{D_2}}
        \left[  \ln\left( \frac{n_2^{-+}}{d_2^{-}}\right)
               -\ln\left( \frac{n_2^{--}}{d_2^{-}}\right) \right]
\right\},
\label{intC2minus}
\eq
with
\bq
C_2=C_2^-=\left(1+\frac{\muno^2-\ieps}{\Ps}\right)\mdue^2-(\mtre^2-\ieps)\,.
\eq
For $N$ and part of $M$ we derive:
\bqa
N(y_2)&=&\ln\left(\frac{-\Ps}{\mtre^2-\ieps}\right)+\ln(y_2),
\nll
\ln\left(\frac{n_2^{-+}}{d_2^{-}}\right)&=&\ln\left(\frac{y_{l-}D_{t2}}{\Ps}\right)
+\ln\left(1-\frac{y}{y_{l-}}\right),
\nll
\ln\left(\frac{n_2^{--}}{d_2^{-}}\right)&=&\ln\left(\frac{y_{l+}D_{t2}(y_{l-}-y_{d2})}{\Ps y_{d2}}\right)
+\ln\left(1-\frac{y}{y_{l+}}\right)-\ln\left(1-\frac{y}{y_{d2}}\right),
\nll
\ln(y_2)&=&\ln\left(\frac{2C_2y_{d2}}{y_{l-}y_{l+}D_{t2}}\right)+\ln\left(1-\frac{y}{y_{d2}}\right)
-\ln\left(1-\frac{y}{y_{l-}}\right)-\ln\left(1-\frac{y}{y_{l+}}\right),
\label{ingC2minus}
\eqa
where
\bqa
y_{d2}&=&-\frac{B_2+t_2^{min}C_2}{D_{t2}C_2}\,,
\nll
y_{l-}&=&+\frac{A-t_2^{min}}{D_{t2}}\,,
\nll
y_{l+}&=&-\frac{A+t_2^{min}}{D_{t2}}\,.
\eqa

Putting the variable substitutions and Eqs.~(\ref{ingC2minus})
into Eq.~(\ref{intC2minus}), one gets:
\bqa
J_2&=&-\frac{1}{2\sqrt{D_{T}}}
\ln\left(\frac{-\Ps}{\mtre^2-\ieps}\right)\ln\left(\frac{1-y_{T2}}{-y_{T2}}\right)+J'_2\,,
\eqa
with
%\nll
%J'_2&=&-\frac{1}{2\sqrt{D_{T}}}\int_0^1dy\biggl\{
%\biggl[
%\ln\left(\frac{2C_2y_{d2}}{y_{l-}y_{l+}D_{t2}}\right)+\ln\left(1-\frac{y}{y_{d2}}\right)
%-\ln\left(1-\frac{y}{y_{l-}}\right)-\ln\left(1-\frac{y}{y_{l+}}\right)
%\biggr]
%\nll&&\times
%\biggl[\frac{1}{y-y_{d2}}-\frac{1}{y-y_{l-}}-\frac{1}{y-y_{l+}}\biggr]
%-
%\biggl[
%\ln\left(\frac{y_{l-}D_{t2}}{\Ps}\right)+\ln\left(1-\frac{y}{y_{l-}}\right)
%\biggr]\frac{1}{y-y_{d2}}
%\nll
%&&+
%\biggl[
%\ln\left(\frac{y_{l+}D_{t2}(y_{l-}-y_{d2})}{\Ps y_{d2}}\right)+\ln\left(1-\frac{y}{y_{l+}}\right)
%-\ln\left(1-\frac{y}{y_{d2}}\right)
%\biggr]\frac{1}{y-y_{d2}}
%\biggr\}\,.
%\eqa
%
%Finally,
\bqa
\hspace*{-5mm}
   J'_2&=&\frac{1}{2(\Ps+\muno^2)}\biggl\{
   -\ln\left(\frac{2C_2y_{d2}}{y_{l-}y_{l+}D_{t2}}\right)\left[l\left(y_{l+}\right)
                                                      +\left(y_{l-}\right)\right]
   +\frac{1}{2}l^2\left(y_{l+}\right)
   +\frac{1}{2}l^2\left(y_{l-}\right)
\nll &&
   +\biggl[\ln\left(\frac{2C_2y_{d2}}{y_{l-}y_{l+}D_{t2}}\right)-\ln\left(\frac{y_{l-}D_{t2}}{\Ps}\right)
   +\ln\left(\frac{y_{l+}D_{t2}(y_{l-}-y_{d2})}{\Ps y_{d2}}\right)
    \biggr]l\left(y_{d2}\right)
\nll &&
   +{\cal M}(y_{l-},y_{l+})-{\cal M}(y_{d2} ,y_{l+})
   +{\cal M}(y_{l+},y_{l-})-{\cal M}(y_{d2} ,y_{l-})
   -2{\cal M}(y_{l-},y_{d2})
\biggr\},
\eqa
and
\bq
l(y)=\ln\left(1-\frac{1}{y}\right),
\eq
and the ``master integral'':
\bq
{\cal {M}}(\yl,\yd)=\int^1_0\frac{ dy}{(y-\yd)}\ln\left(1-\frac{y}{\yl}\right)
= \ln\left(1-\frac{\yd}{\yl}\right) l(\yd)
          -\Litwo\left(\frac{1-\yd}{\yl-\yd}\right)
          +\Litwo\left(\frac{ -\yd}{\yl-\yd}\right).
\label{master}
\eq

$\bullet$ {Solution $C_2^+$}
%---------------------------
\bq
I_2(y_2)=
-\frac{1}{2\sqrt{D_{T}}y_2}\left\{
N(y_2)+\frac{C_2^+}{\sqrt{D_2}}
        \left[  \ln\left( \frac{n_2^{++}}{d_2^{+}}\right)
               -\ln\left( \frac{n_2^{+-}}{d_2^{+}}\right) \right]
\right\},
\label{intC2plus}
\eq
with
\bq
C_2=C_2^+=-\left(1+\frac{\muno^2-\ieps}{\Ps}\right)\mdue^2+(\mtre^2-\ieps)\,.
\eq

The derivation is similar and we limit ourselves to presenting the answer for $J'_2$:
%\bqa
%J'_2&=&-\frac{1}{2\sqrt{D_{T}}}\int_0^1dy\biggl\{
%\biggl[
%\ln\left(\frac{2C_2y_{d2}}{y_{l-}y_{l+}D_{t2}}\right)+\ln\left(1-\frac{y}{y_{d2}}\right)
%-\ln\left(1-\frac{y}{y_{l-}}\right)-\ln\left(1-\frac{y}{y_{l+}}\right)
%\biggr]
%\nll&&\times
%\biggl[\frac{1}{y-y_{d2}}-\frac{1}{y-y_{l-}}-\frac{1}{y-y_{l+}}\biggr]
%+
%\biggl[
%\ln\left(\frac{-y_{l+}D_{t2}}{\Ps}\right)+\ln\left(1-\frac{y}{y_{l+}}\right)
%\biggr]\frac{1}{y-y_{d2}}
%\nll
%&&-
%\biggl[
%\ln\left(\frac{-y_{l-}D_{t2}(y_{l+}-y_{d2})}{\Ps y_{d2}}\right)+\ln\left(1-\frac{y}{y_{l-}}\right)
%-\ln\left(1-\frac{y}{y_{d2}}\right)
%\biggr]\frac{1}{y-y_{d2}}
%\biggr\}\,,
%\eqa
%and the final result:
\bqa
\hspace*{-5mm}
 J'_2&=&\frac{1}{2(\Ps+\muno^2)}\biggl\{
   -\ln\left(\frac{2C_2y_{d2}}{y_{l-}y_{l+}D_{t2}}\right)\left[l\left(y_{l+}\right)
                                                      +\left(y_{l-}\right)\right]
   +\frac{1}{2}l^2\left(y_{l+}\right)
   +\frac{1}{2}l^2\left(y_{l-}\right)
\nll &&
   +\biggl[\ln\left(\frac{2C_2y_{d2}}{y_{l-}y_{l+}D_{t2}}\right)-\ln\left(-\frac{y_{l+}D_{t2}}{\Ps}\right)
   +\ln\left(\frac{y_{l-}D_{t2}(y_{d2}-y_{l+})}{\Ps y_{d2}}\right)
    \biggl]l\left(y_{d2}\right)
\nll &&
   +{\cal M}(y_{l-},y_{l+})-{\cal M}(y_{d2} ,y_{l+})
   +{\cal M}(y_{l+},y_{l-})-{\cal M}(y_{d2} ,y_{l-})
   -2{\cal M}(y_{l+},y_{d2})
\biggr\}.
\eqa

\subsection{Definitions of functions $J^{tbud}_{sub}$}
%----------------------------------------------
The mass singularities in arguments of the logarithms may be compensated
by combining $J$ with one more $C_0$ function:
\bqa
\label{Jsub1}
 J^{tbud}_{sub}(\Qs,\Ps;\mbt,\mtp,\mw)  &=&
 J(\Qs,\Ps;\mbt,\mtp,\mdn,\mup,\mw)
\nll &&
 -\left(1+\frac{\Qs}{\mbtll+\Ps}\right) C_0(-\mupll,-\mdnll,\Qs;\mw,\mdn,0)\,,
\\
 J^{tbud}_{sub}(\Qs,\Ps;\mtp,\mbt,\mw)  &=&
 J(\Qs,\Ps;\mtp,\mbt,\mup,\mdn,\mw)
\nll &&
-\left(1+\frac{\Qs}{\mtpll+\Ps}\right) C_0(-\mdnll,-\mupll,\Qs;\mw,\mup,0)\,.
\label{Jsub2}
\eqa
where $\Ps=\Us$ for cross functions.
The two mass-singular $C_0$ functions appearing in Eqs.~(\ref{Jsub1}) and (\ref{Jsub2}) cancel
in the total expression for the EW correction which proves the absence in it of logarithmic mass
singularities (not KLN theorem!).

\subsubsection{Calculation of the subtracted function $J^{tbud}_{sub}$}
%---------------------------------------------------------------
The first function $J^{tbud}_{sub}$ is given by equation (\ref{Jsub1}),
where we may neglect $\mup$ which does not give rise to a mass singularity:
\bq
J^{tbud}_{sub}=J - \left(1+\frac{\Qs}{\Ps+\muno^2}\right) C_0(0,\mqat^2,\Qs;\mtre,\mqat,0)\,.
\eq
For $C_0$ we have:
\bq
C_0(0,\mqat^2,\Qs;\mtre,\mqat,0)=\frac{1}{\Qs}\left[\ln\left(-\frac{\Qs}{\mqat^2}\right)\ln(R_3)
     +\Litwo(R_3)-\zeta(2)\right]\,,
\eq
with
\bq
R_3 = \frac{\Qs+\mtre^2-\ieps}{\mtre^2}\,.
\label{R3}
\eq
For $J^{tbud}_{sub}$ one derives an expression for $J_0$ that is free of the $\mqat$ mass singularity;
$J_{1,2}$ remain unchanged:
\bq
\label{Jsub}
J^{tbud}_{sub} = J^{tbud}_{sub,0}+J_1+J_2\,,
\eq
\bq
J^{tbud}_{sub,0}=
     -\frac{1}{\Ps+\muno^2}\biggl\{
       \ln(R_3)\biggl[\ln\left(-\frac{\Qs}{\mtre^2}\right)-\ln\left(R_{13}\right)
      -\ln\left(R_{11}\right)\biggr]
      +\Litwo(R_3)-\Litwo\left(1-R_{13}\right)\biggl\}\,.
\label{Jsub0}
\eq
%--

\subsection{Definitions of the function $J^{tbud}_{subsub}$}
%-------------------------------------------------------
 If we want to neglect the $b$ quark mass, $\mbt$, we must perform a second subtraction of the mass 
singular $C_0$ function $C_0(-\mtpll,-\mbtll,\Qs,\mw,\mbt,0)$ that appears in the limit $\mbt\to 0$.

 Note that only one of $J^{tbud}_{sub}$ contains an $\mbt$ mass singularity.
\bqa
J^{tbud}_{subsub,1}(\Qs,\Ps;\mbt,\mtp,\mw) &=& J^{tbud}_{sub}(\Qs,\Ps;\mbt,\mtp,\mw)
\nll
&&  -\frac{\Qs+\mtpll}{\Ps} C_0(-\mtpll,-\mbtll,\Qs;\mw,\mbt,0).
\label{subsub1}
\eqa
 Again, the $\mbt$ mass singular $C_0$ function
$C_0(-\mtpll,-\mbtll,\Qs;\mw,\mbt,0)$ cancels in the total EW correction.

 Since we do not want to consider the limit $\mtp=0$, we simply rename the second function:
\bqa
J^{tbud}_{subsub,2}(\Qs,\Ps;\mtp,\mbt,\mw) = J^{tbud}_{sub}(\Qs,\Ps;\mtp,\mbt,\mw),
\eqa
assuming $\mbt=0$ for this, non-singular case.

\subsubsection{The first function $J^{tbud}_{sub}$ in the limit $\mbt\to 0$}
%---------------------------------------------------------------------
Here we simply present the limits of $J_{0,2}$ which develop $\muno$ mass
singular logarithms; for $J_{1}$ it is sufficient to set $\muno=0$ in Eq.~(\ref{J1ala2006}).
\bqa
J_0 &=& \frac{1}{\Ps} \biggl\{\ln\left( \frac{P}{\mtre^2}\right)
  \left[ 2\ln\left(\frac{\Ps-\ieps}{\mtre^2}\right)+\ln\left(\frac{\mtre^2}{\muno^2}\right)\right]
 -\frac{1}{2} \ln\left(-\frac{\Ps}{\mtre^2}\right) \ln\left(\frac{\Ps-\ieps}{\muno^2}\right)
\nll &&
     -\ln\left(\frac{P}{\mtre^2}\right) \ln\left(-\frac{\Qs}{\mtre^2-\ieps}\right)
     -\Litwo\left( R_3 \right)
     +\Litwo\left(-\frac{\Ps-\mtre^2-\ieps}{\mtre^2}\right)
                    \biggr\}.
\label{J0lim1}
\eqa
Here $P$ is given by Eq.~(\ref{Predefined}).
For the $J_2$ part one finds:
\bqa
J_2 &=& \frac{1}{\Ps}
\left[\ln\left(\frac{\Delta_{23}+\ieps}{\Ps} \right)
     +\frac{1}{2}\ln\left(-\frac{\Ps}{\mtre^2}\right)\right] \ln\left(\frac{\muno^2}{\mtre^2}\right)
\nll &&
%%% +\frac{1}{2\Ps}\biggl[
%%%         -\ln\left(-\frac{\Ps-\ieps}{\Ps}\right) l\left(\zlp\right)
%%%         +\frac{1}{2} l^2\left(\zlp\right)
%%%         -\ln\left(-\frac{\Ps-\ieps}{\Ps}\right) l\left(\zlm\right)
%%%         +\frac{1}{2} l^2\left(\zlm \right)
 +\frac{1}{2\Ps}\biggl\{
          \frac{1}{2} l^2\left(\zlp\right)
         +i\pi l\left(\zlp\right)
         +\frac{1}{2} l^2\left(\zlm \right)
         +i\pi l\left(\zlm\right)
\nll &&
       +\left[ 2 \ln\left( \frac{\Delta_{23}+\ieps}{\Ps}\right)+\ln\left(-\frac{\Ps}{\mtre^2}\right)
\right]
                       \ln\left(-\frac{\mtre^2 \mdue^2}{\tilde\Delta_{23}^2}\right)
\nll &&
         +{\cal M}(\zlm,\zlp)
         -{\cal M}(1,\zlp)
         +{\cal M}(\zlp,\zlm)
         -{\cal M}(1,\zlm)
       -2 \Litwo\left(\frac{1}{1-\zlm}\right)
                  \biggr\},
\label{J2lim1}
\eqa
where
\bq
\zlm = \frac{\Ps}{(\Ps-\ieps)}\frac{\tilde\Delta_{23}}{(\tilde\Delta_{23}+\Ps)}\,,
\qquad
\zlp = -\frac{\tilde\Delta_{23}}{\mtre^2}\,,
\eq
and
\bq
\tilde\Delta_{23}=\mdue^2-(\mtre^2-\ieps).
\label{tildeDelta}
\eq
Summing Eqs.~(\ref{J0lim1}) and (\ref{J2lim1}) with Eq.~(\ref{J1ala2006})
in the limit of $\muno=0$ we get:
\bq
J^{tbud}_{sub}=J_0+J_1+J_2.
\eq
%--

\subsubsection{The function $J^{tbud}_{subsub,1}$}
%------------------------------------------
The $C_0$ function needed in Eq.~(\ref{subsub1}) with explicitly separated out $\mdue$
mass singular logarithm looks as follows:
% p12=-mtp2, p22=-mbt2, m12=mw2, m22=mbp2
\bqa
&&C_0(p_1^2,p_2^2,\Qs;\muno^2,\mdue^2,0)=\frac{1}{\Qs-p_1^2}\biggl\{
\ln\left(\frac{\muno^2}{\mdue^2}\right)\ln\left(\frac{\P}{p_1^2+\muno^2-\ieps}\right)
\nll &&
 +\left[\ln\left(-\frac{(\Qs-p_1^2)^2}{\muno^2\Qs}\right)-2i\pi\right]
        \ln\left(\frac{\P}{\muno^2}\right)
 -\left[\ln\left(-\frac{(\Qs-p_1^2)^2}{\muno^2p_1^2}\right)-2i\pi\right]
        \ln\left(\frac{p_1^2+\muno^2-\ieps}{\muno^2}\right)
\nll &&
 -\Litwo\left(\frac{\P}{\muno^2}\right)
 +\Litwo\left(\frac{p_1^2+\muno^2-\ieps}{\muno^2}\right)
 +\Litwo\left(\frac{\P}{p_1^2+\muno^2}\right)
 +\Litwo\left(\frac{(\P)^*}{p_1^2+\muno^2}\right)
 -2\zeta(2) \biggr\}.
\label{C0forsub1}
\eqa
Here $\P=\Qs+\muno^2-\ieps$ and $p_2^2=-\mdue^2\,$.

Let us emphasize that in this section $C_0$ has its own list of dummy arguments!

In order to derive $J^{tbud}_{subsub,1}$ it is sufficient to redefine $J_0$ of~(\ref{J0lim1})
into $J'_0$ by summing it with the first row of $J_2$ of~(\ref{J2lim1})
and the $C_0$ function of~(\ref{C0forsub1}) with the coefficient of Eq.~(\ref{Jsub1}),
i.e. collect together all $\mbt$ mass singular terms; that is:
\bq
J^{tbud}_{subsub,1}=J'_0+J_1+J'_2,
\label{Jsubsub1}
\eq
%--
where $J_1$ remains unchanged, given as before by the limit of Eq.~(\ref{J1ala2006})
at $\muno=0$, and $J'_2$ denotes the rest of $J_2$ without its first row.

For $J'_0$ we get:
\bqa
J'_0&=&\frac{1}{\Ps}\biggl\{
\ln\left(-\frac{\Ps-\ieps}{\Qs+\mdue^2}\right)\ln\left(\frac{\P}{\mtre^2}\right)
+\left[\ln\left(-\frac{\Qs+\mdue^2}{\mtre^2-\ieps}\right)+\ln\left(\frac{\mtre^2}{\mdue^2}\right)
 \right]
 \ln\left(-\frac{\tilde\Delta_{23}}{\mtre^2}\right)
\nll &&
-\frac{1}{2}\ln\left(-\frac{\Ps-\ieps}{\mtre^2}\right)\ln\left(\frac{\Ps-\ieps}{\mtre^2}\right)
 +\Litwo\left(-\frac{\Ps-\mtre^2-\ieps}{\mtre^2}\right)
\nll &&
 -\Litwo\left(-\frac{\tilde\Delta_{23}}{\mtre^2}\right)
 -\Litwo\left(-\frac{\P}{\Delta_{23}}\right)
 -\Litwo\left(-\frac{(\P)^*}{\Delta_{23}}\right)
 +2\zeta(2) \biggr\},
\label{Jsubsub10}
\eqa
%---
with $\tilde\Delta_{23}$ defined by Eq.~(\ref{tildeDelta}).

In Table~\ref{table1} we give a comparison of real and imaginary parts of the function $J^{tbud}_{subsub,1}$
defined by Eq.~(\ref{subsub1}) and related ones, computed with the aid of the
LoopTools package~\cite{Hahn:1998yk},
vs numbers derived exactly from Eq.~(\ref{Jsubsub1}) with
$J'_0$ given by Eq.~(\ref{Jsubsub10}) and $J_1$ and $J'_2$ by previous equations as explained 
just below Eq.~(\ref{Jsubsub1}), i.e. results of this paper.

 The numbers are given for four values of $s$ (the first two values near the
kinematical edges and the latter two in the $\pm$1 GeV vicinity of the $W$
resonance) and for three values of $\cos\theta$
at $\mgm=10^{-40},~\muno=10~^{-7},~\mdue=174.3,~\mtre=80.403,~\mqat=5\cdot
10^{-7}$ (all masses are given in GeV);
first lines  --- LoopTools, second lines --- this paper.
%\vspace*{-6mm}

%/Export_JAW_tbln_11_08_2008_r16_paper
%Make_JAWsubsub_mb1
%main_JAWsubsub_mb1.F
%     JAWsubsub_mb1.f
%main_JAWsubsub_mb1.log_for_paper_fin

\begin{table}[!hb]
{\small
\hspace*{-5mm}
\caption{Comparison of $J^{tbud}_{subsub,1}\label{table1}$}
\begin{tabular}{||c|c|c||}
\hline
\hline
 $\cos\theta$  & $\sqrt{s}=1$GeV               & $\sqrt{s}=173.2$GeV                  \\
\hline
-0.999 &-0.374523885975E-7, 0.483905777792E-7 & 0.220728294741E-1, -0.404470353912E-1 \\
%               |6                      |11                  |12                 |12
       &-0.374523675918E-7, 0.483905777797E-7 & 0.220728294741E-1, -0.404470353912E-1\\
\hline
0      &-0.103142713149E-3, 0.143516783301E-3 & 0.490255707173E-1, -0.933810555325E-1\\
%                   |11                  |12                 |12                 |12
       &-0.103142713128E-3, 0.143516783301E-3 & 0.490255707173E-1, -0.933810555325E-1\\
\hline
0.999  &-0.779612000864,     1.57384208867     & 99.5783633460,     -218.316883034    \\
%                     |12                |12               |10                   |12
       &-0.779612000864,     1.57384208867     & 99.5783633453,     -218.316883034    \\
\hline
\hline
               & $\sqrt{s}=79.403$GeV          & $\sqrt{s}=81.403$GeV                 \\
\hline
-0.999 & 0.227686132083E-4,   -0.303846732065E-4 & 0.243884328473E-4, -0.327139389502E-4\\
%                   |10                    |12                |10                   |12
       & 0.227686132074E-4,   -0.303846732065E-4 & 0.243884328481E-4, -0.327139389502E-4\\
\hline
0      &-0.504773282953E-3,   ~0.120287840705E-3& -0.507195719751E-3, -0.248876535390E-3\\
%                    |11                    |12               |10                   |12
       &-0.504773282955E-3,   ~0.120287840705E-3& -0.507195719749E-3, -0.248876535390E-3\\
\hline
0.999  &-5.59686285604,       ~1.92566453330     &-5.65305814019,     -2.07873259553    \\
%                    |12                   |12                 |12                 |12
       &-5.59686285604,       ~1.92566453330     &-5.65305814019,     -2.07873259553    \\
\hline
\hline
\end{tabular}
}
\end{table}
%\vspace*{-2mm}

\noindent
As is seen from the Table, there is agreement within 9-12 digits for real
parts and within 12 digits for imaginary parts, which seems quite
satisfactory given that we are using only Double Precision in the Fortran code.
We also note that exact formulae with extremely small values of masses
$\mgm,~\muno,~\mqat$ together with large masses $\mdue,~\mtre$  occur
in LoopTools which could result in loss of computational precision.
Unlike LoopTools, the formulae derived in this paper are rather compact
and do not explicitly contain the masses $\mgm,~\muno,~\mqat$; the latter property
is the main goal of this paper. Also, the short formulae of this paper
demonstrate the underlying physics of singularity cancellations and,
moreover, their execution is much faster than LoopTools.
On the other hand, one may also conclude that this comparison proves
the high reliability  of the LoopTools package.

\subsubsection{The function $J^{tbud}_{subsub,2}$}
%-----------------------------------------
This function is derived by means of simply setting $\mdue=0$ in Eq.~(\ref{Jsub}). Let,
\bq
J^{tbud}_{subsub,2}=J^0_{0}+J^0_1+J^0_2\,.
\label{subsub2}
\eq
Since $J^{tbud}_{sub,0}$ is independent of $\mdue$, we have
\bq
J^0_{0}=J^{tbud}_{sub,0}\,,
\eq
with the latter given by Eq.~(\ref{Jsub})

For the part $J^0_1$ one gets:
\bqa
J^0_{1} &=& \ln\left(R_{13}\right)\ln\left(\frac{\Ps+\Delta_{13}+\ieps}{\Delta_{13}}\right)
\nll &&
         +\Litwo\left(\frac{\Ps}{\Ps+\muno^2} \right)
         -\Litwo\left[\left(-\frac{\Ps}{\Delta_{13}}-\ieps\right) \frac{1}{R_{13}} \right]
         +\Litwo\left( -\frac{\Ps}{\Delta_{13}}-\ieps \right)\,.
\eqa

For the second part, $J^0_{2}$, we introduce some common notation:
\bqa
 t^{\max} &=& \frac{1}{y^{\max}}\left[ \sqrt{T^4 (y^{\max})^2
             +2 \Ps y^{\max} \sqrt{C_2} +C_2}-\sqrt{C_2} \right],
\nll
y^{\max} &=&  -\frac{\muno^2-\ieps}{\Ps}\,,
\qquad
y^{\min}\;=\; -1-\frac{\muno^2-\ieps}{\Ps}\,,
\nll
 \sqrt{C_2} &=& \mtre^2-\ieps.
\nll
\eqa

There are two solutions:\\

$\bullet$ if  $\Ps+\muno^2-\mtre^2 \geq 0$, then
\bqa
J^0_{2} &=&\frac{1}{2(\Ps+\muno^2)}\Biggl\{
\left[ \ln\left(\zlm \frac{\dt}{\Ps}\left(1-\frac{\zlp}{\zlm}\right)\right)
              -\ln\left(-\zlp \frac{\dt}{\Ps}\right)
              -\ln\left(1-\frac{\zlm}{\zlp}\right)\right] l\left( \zlm \right)
\nll &&
            +\frac{1}{2} l^2\left(\zlp\right)
            -\ln\left(\frac{2 \sqrt{C_2}}{\zlp\dt}\right) l\left(\zlp\right)
            +\Litwo\left(\frac{1-\zlm}{\zlp-\zlm}\right)
            -\Litwo\left(\frac{ -\zlm}{\zlp-\zlm}\right)   \Biggr\},
\eqa
where
\bqa
t^{\min} &=& -\Ps-\frac{2 (\mtre^2-\ieps)}{y^{\min}}\,,
\nll
\dt &=& t^{\max}-t^{\min},
\nll
{y_{l\mp}} &=& \pm\frac{\Ps\mp t^{\min}}{\dt}\,.
\eqa

$\bullet$ if $\Ps+\muno^2-\mtre^2 \leq 0$, then

\bqa
J^0_{2} &=&\frac{1}{2(\Ps+\muno^2)}\Biggl\{
           \frac{1}{2}\ln^2\left(\frac{\dt}{2\Ps}\right)
          -\frac{1}{2}\ln^2\left(-\frac{\dt \sqrt{C_2}}{2\Ps (y^{\min}\Ps+\sqrt{C_2})}\right)
\nll &&
          +\ln\left(\frac{2\sqrt{C_2}}{\zlp\dt}\right)\ln\left(-\frac{\dt \sqrt{C_2}}{y^{\min}(\Ps)^2}\right)
          -\ln\left(-\frac{2 y^{\min}\Ps}{\dt\zlp}\right) \ln\left(\frac{\dt}{2\Ps}\right)
\nll &&
          -\ln\left(1+\frac{\sqrt{C_2}}{y^{\min}\Ps}\right) \ln\left(\frac{\dt}{2\Ps}\right)
          -\ln\left(\frac{2 \sqrt{C_2}}{\zlp\dt}\right) l\left( \zlp \right)
\nll &&
          +\frac{1}{2} l^2\left( \zlp\right)
          +\Litwo\left(\frac{1}{\zlp}\right)
          -\Litwo\left(\frac{\sqrt{C_2}}{y^{\min} \Ps+\sqrt{C_2}}\right)-\zeta(2)\Biggr\},
\eqa
with
\bqa
t^{\min} &=& \Ps,
\nll
\dt  &=& t^{\max}-t^{\min},
\nll
\zlp &=& -\frac{2 \Ps}{\dt}.
\eqa

In Table~\ref{table2} we give a similar comparison as in Table~\ref{table1} but now it is for the function
$J^{tbud}_{subsub,2}$ defined by Eq.~(\ref{subsub2}) and related ones. The setup is the same as for Table~\ref{table1}.
Again, first lines  --- LoopTools, second lines --- this paper.
%\vspace*{-3mm}

%/Export_JAW_tbln_11_08_2008_r16_paper
%Make_JAWsubsub_mb2
%main_JAWsubsub_mb2.F
%     JAWsubsub_mb2.f
%main_JAWsubsub_mb2.log_for_paper_fin

\begin{table}[!hb]
\centering
{\small
\caption{Comparison of $J^{tbud}_{subsub,2}\label{table2}$}
\begin{tabular}{||c|c|c||}
\hline
\hline
 $\cos\theta$  & $\sqrt{s}=1$GeV               & $\sqrt{s}=173.2$GeV                  \\
\hline
-0.999 &-0.943679024194E-3,   0                &-0.495811769871E-6,  0.120460853637E-5\\
%                    |11                                    |10                   |12
       &-0.943679024191E-3,   0                &-0.495811769847E-6,  0.120460853637E-5\\
\hline
0      &-0.177753225378E-3,   0                &-0.373359924505E-6, -0.345101621985E-8\\
%                   |10                                    |9                   |10
       &-0.177753225357E-3,   0                &-0.373359924465E-6, -0.345101621921E-8\\
\hline
0.999  &-0.128202726721E-3,   0                &-0.253792663608E-6, -0.119084548438E-5\\
%                   |10                                      |11                  |12
SANC=  &-0.128202726700E-3    0                &-0.253792663600E-6, -0.119084548438E-5\\
\hline
\hline
               & $\sqrt{s}=79.403$GeV          & $\sqrt{s}=81.403$GeV                 \\
\hline
-0.999 & 0.917891464851E-3,   0                & 0.843008928372E-3,  0.718400159039E-3\\
%                   |10                                      |11                  |12
       & 0.917891464846E-3,   0                & 0.843008928376E-3,  0.718400159039E-3\\
\hline
0      &-0.188317064868E-3,   0                &-0.177932415951E-3, -0.901230697461E-4\\
%                    |11                                     |11                  |12
       &-0.188317064870E-3,   0                &-0.177932415950E-3, -0.901230697461E-4\\
\hline
0.999  &-0.266362647090E-3,   0                &-0.258030469532E-3, -0.157505612576E-3\\
%                    |11                                     |11                  |12
       &-0.266362647091E-3,   0                &-0.258030469531E-3, -0.157505612576E-3\\
\hline
\hline
\end{tabular}
}
\end{table}
%%\vspace*{-2mm}
%\vspace*{2mm}

\noindent
As seen from the Table, there is again agreement within 10-12 digits for real and imaginary parts.
Below the $W$ resonance this function is real.

We switch now to another example of $J$ functions.

\section{Calculation of $J$ for the process $ud \rightarrow tb$\label{s-channel}}
%--------------------------------------------------------------------------------
As in the previous section we shall consider only the direct and $\gamma W$ diagrams.
For the process $ud \rightarrow tb$ the list of arguments of the universal
function $J$ is: $m_1^2=\mbtll$, $m_2^2=\mtpll$, $m_3^2=\mwll$, $m_4^2=m_q$ --- singular mass.
\bqa
J&\equiv &J_{AW}^d(\Qs,\Ps;\mdn,\mup,\mbt,\mtp)
\eqa
%      + d0( - mdn^2, - mup^2, - mtp^2, - mbt^2,Qs,Ts,1,0,1,mdn,1,mw,1,mbt)* ( Qs + mw^2 )
%      + c0( - mtp^2, - mbt^2,Qs,1,mw,1,mbt,1,0)
%      - c0( - mbt^2, - mdn^2,Ts,1,mbt,1,0,1,mdn)
In the following presentation we change the notation of masses as in (\ref{mass_redenote}) and
neglect the mass $\mup$ which does not lead to a mass singularity.

\subsection{Integration with respect to $z,x$}
%---------------------------------------------
We omit the details of the integrations with respect to $z$ and to $x$.
The standard ingredients are:
\bqa
N_{xy}&=&  -2 k_{xy} p_2 = N_{y}\,,
\nll
N_{y} &=& \Ps(1-y) + \Qs + \mdue^2(1-y) + \mqat^2
\nll
k^2_{xy}&=& N_y x - T^2_y\,,
\nll
T^2_y&=& \Ps y(1-y)+\muno^2(1-y)+\mqat^2y\,,
\nll
L &=& \P x - \ieps\,.
\label{ingudtb1}
\eqa
Contrary to the case of $t\to bud$ decay, the variables $k^2_{xy}$ and
$L^*=L-k^2_{xy}$ are linear in $x$ in this case since we have neglected $\mup$:
\bqa
L^*&=& L-k^2_{xy}=[-(\Ps+\mdue^2)(1-y)+\mtre^2-\mqat^2]x+T^2_y-\ieps\,.
\label{ingudtb2}
\eqa

\subsection{Integration over $y$}
%--------------------------------
The one-dimensional integral with respect to $y\to (1-y)$ is
\begin{equation}
   J = \int\limits_{0}^{1}\,dy I(y).
\label{Judtb}
\end{equation}
With the aid of Eq.~(\ref{intx}) and with ingredients of Eqs.~(\ref{ingudtb1}--\ref{ingudtb2})
for the integrand $I(y)$ one obtains a simpler result than
Eqs.~(\ref{IntegrOvery}--\ref{Ipofy}):
\bqa
I(y)=\left( -\frac{1}{k^2_{{xy}|y}} - \frac{1}{T^2_y-\ieps} \right)
             \left[\ln(L^*{_|{_{_y}}})-\ln(\P y)\right].
\eqa
where
\bqa
\P &=& \Qs+\mtre^2-\ieps\,,
\nll
T^2_y&=&\Ps y (1-y)+\muno^2y+\mqat^2(1-y)\,,
\nll
k^2_{{xy}|y}   &=&  \mdue^2 y(1-y)-\muno^2y+\Qs(1-y)\,,
\nll
L^*{_|{_{_y}}} &=& -\mdue^2 y(1-y)+\muno^2y+\mtre^2(1-y)-\ieps\,.
\eqa
%---

\subsubsection{Integrand in the limit $\mqat\to 0$}
%--------------------------------------------------
After some algebra one can derive the following expression for $I(y)$ valid
in the limit $\mqat\to 0$:
\bqa
 I(y)=  \frac{1}{\sdkxy}     \Bigl[{\cal T}(y,\ykxyl)-{\cal T}(y,\ykxyll)\Bigr]
       + \frac{1}{(\Ps+\muno^2)} \Bigl[{\cal T}(y,\yPslL)-{\cal T}(y,\yPsllL)\Bigr].
\eqa
Thus $I(y)$ is expressed in terms of the auxiliary function
\bqa
{\cal T}(y,\eta) = \frac{1}{(y-\eta)}
                  \biggl[
                    \ln\left(1-\frac{y}{\yLmkl} \right)
                   +\ln\left(1-\frac{y}{\yLmkll}\right)
                   -\ln(1-y)
                   -\ln\left(\frac{\P}{\mtre^2}\right) \biggr]
\eqa
with roots of the quadratic trinomial ${T^2_y-\ieps}$:
\bqa
 \yPslL = \frac{\Ps+\muno^2-\ieps}{\Ps}\,, \qquad
 \yPsllL=-\frac{\mqat^2-\ieps}{\Ps+\muno^2-\ieps}\,.
\eqa
%* $\yPsllL$ is not needed.
%* m1=mbt, m2=mtp, m3=mw, m4=mq-singular
%*
The other objects arise from the trinomial $k^2_{{xy}|y}$ in $y$
\bqa
k^2_{{xy}|y}&=&-\mdue^2 y^2+(\Delta_{21}-\Qs) y+(\Qs+\ieps), \nll
\bkxyy &=& \Delta_{21}-\Qs,     \nll
\sdkxy &=&\sqrt{\bkxyy^2+4 \mdue^2 (\Qs+\ieps)}
\eqa
with roots:
\bqa
\ykxyl = \frac{-\bkxyy+\sdkxy}{2(-\mdue^2)}\,,\qquad
\ykxyll= \frac{-\bkxyy-\sdkxy}{2(-\mdue^2)}\,,
\eqa
and trinomial $L^*{_|{_{_y}}}$ in $y$, i.e.
\bqa
% Lmk2dy=argLmk=y^2*mtp^2+y*(-mtp^2+mbt^2-mw^2)+mw^2
L^*{_|{_{_y}}} &=& \mdue^2 y^2+y (\Delta_{12}-\mtre^2)+(\mtre^2-\ieps), \nll
\bLmk  &=& \Delta_{12}-\mtre^2, \nll
\sdLmk &=& \sqrt{\bLmk^2-4 \mdue^2(\mtre^2-\ieps)}
\eqa
with roots:
\bqa
\yLmkl  = \frac{-\bLmk+\sdLmk}{ 2 \mdue^2}\,, \qquad
\yLmkll = \frac{-\bLmk-\sdLmk}{ 2 \mdue^2}\,.
\eqa
%---

\subsection{Function $J$}
%------------------------
The final answer for the integral $J$ of Eq.~(\ref{Judtb}) is split into two parts,
\bqa
J = J_{0}+J_{add}.
\eqa
The first part, free of $\mqat^2$ mass singularity, is
\bqa
J_{0} &=& \frac{1}{\sdkxy} \biggl[ \bigg\{
\left[\ln(1-\ykxyl)+\ln(R_3) \right] l(\ykxyl)
          -{\cal{M}}(\yLmkl,\ykxylR)-{\cal{M}}(\yLmkll,\ykxylR)
\nll &&
         +\Litwo\left(-\frac{\ykxyl}{1-\ykxyl}\right)
\bigg\}
           - \bigg\{\ykxyl \rightarrow \ykxyll \bigg\}
\Biggr]
\nll &&
           +\frac{1}{\Ps+\muno^2}
\Biggl[
             \left\{ {\cal{M}}(\yLmkl,y_1) + \Litwo\left(\frac{1}{\yLmkl}\right)\right\}
           + \left\{ \yLmkl \rightarrow \yLmkll \right\}
\nll &&
   -\left[ \ln(1-\yPslL)+\ln(R_3) \right] l(\yPslL)
           -\Litwo\left(-\frac{\yPslL}{1-\yPslL}\right)
\biggr],
\label{J0udtb}
\eqa
and the additional part that depends on $\mqat^2$:
\bqa
J^{add}= \frac{1}{\Ps+\muno^2} \ln (R_3) \ln\left( R_{14}\right),
\eqa
For simplicity we introduce the notation
\bq
R_{14}= \frac{\Ps+\muno^2-\ieps}{\mqat^2}\,.
\eq
%--

\subsection{Subtracted function $J^{udtb}_{sub}$}
%-----------------------------------------
The {\it subtracted} function $J^{udtb}_{sub}$ is defined by the following equation
(see Ref.~\cite{Bardin:2009wv}):
\bqa
J^{udtb}_{sub}= J - \frac{\Qs}{\Ps+\muno^2}\, C_0(0,\mqat^2,\Qs;\mtre,\mqat,0)\,.
\eqa
After adding to this the mass singular $C_0$ function,
we get the following expression free of $\mqat$ mass singularity:
\bqa
 J^{udtb}_{sub} &=& J_0
           +\ln(R_3)\left[\ln\left(-\frac{\Qs}{\mtre^2}\right)+\ln\left(R_{13}\right)\right]
           -\Litwo(R_3)+\zeta(2)\,,
\label{Jsubudtb}
\eqa
where $R_{3}$ and $R_{13}$ are defined by Eqs.~(\ref{R3},\ref{R1i}) respectively.
%    & -QLOG(-Qs/rm32)*CQLOG((Qs+m32)/rm32)
%    & -XSPENZ((Qs+m32)/rm32)
%    & +zet2
%      =
%    & +1d0/2*(LOG(Qs/(Qs+m32-ieps)))**2
%    & -1d0/2*(LOG(Qs/(m32-ieps)))**2
%    & +XSPENZ(rm32/(Qs+m32))
%    & -zet2

\subsubsection{Second substraction $J^{udtb}_{subsub,1}$}
%------------------------------------------------
A second substraction eliminates the $b$ quark mass singularity:
%*     m1=mbt, m2=mtp, m3=mw, m4=mq-singular
%       JAW=JAWstpssub2(Qs,Ps,mbt,mtp,mw)
\bqa
\label{Jsubsub1_udtb_def}
J^{udtb}_{subsub,1}(\Qs,\Ps;\mbt,\mtp,\mw) &=&  J^{udtb}_{sub,1}(\Qs,\Ps;\mbt,\mtp,\mw)
\nll &&
%     & +Qs/(Ps+m12)*c0(0d0,m42,-Qs,m32,m42,0d0)
%     & +(Qs+m22+Ps)/Ps*c0(m22,m12,-Qs,m32,m12,0d0)
      -\frac{(\Qs+\Ps+\mtp^2)}{\Ps} C_0(-\mtp^2,-\mbt^2,\Qs;\mw,\mbt,0)\,.\quad
\eqa
The limit in $\mbt$ exists and reads
\bqa
\label{Jsubsub1_udtb_answ}
J^{udtb}_{subsub,1} &=&  \frac{1}{\Ps} \biggl[
      -\ln\left( R_3 \right) \ln\left(\frac{P}{\Ps}\right)
      +\ln\left(-\frac{P}{\Delta_{23}} \right)
       \ln\left(\frac{(\Qs+m_3^2)^2}{(\Qs+m_2^2)^2}\frac{\Ps}{m_3^2}\right)
\\ &&
      +2 \Litwo\left(-\frac{\Delta_{23}}{P}\right)
        +\Litwo\left(1+r_{23}\right)
        -\Litwo\left(\frac{m_2^2}{\Delta_{23}+\ieps}\right)
        -3 \zeta(2)       \biggr],
\nonumber
\eqa
with
\bq
r_{i3}=\frac{\ds \Ps}{\ds m_i^2-m_3^2+\ieps}\,.
\eq
See in this case Eq.~(\ref{mass_redenote}) for the meaning of $m_i$.

In Table~\ref{table3} we give the comparison of real and imaginary parts of the function $J^{udtb}_{subsub,1}$
defined by Eq.~(\ref{Jsubsub1_udtb_def}) and related ones, computed using the LoopTools package with 
numbers, derived exactly from Eq.~(\ref{Jsubsub1_udtb_answ}), i.e. with results of this paper.
 The numbers are given for two values
of variable $s$ (near threshold and at a high value), for three values of $\cos\theta$.
 The masses are
 $\mgm=10^{-40},~\muno=10^{-8},~\mdue=174.3,~\mtre=80.403,~\mqat=10^{-7}$ (all masses are given in GeV);
first lines  --- LoopTools, second lines --- this paper.
%\vspace*{-6mm}

%/Export_JAW_udtb_05_12_2008_r8_paper
%Make_JAW_universal_udtb_subsub_1
%JAW_universal_udtb_subsub_1.F
%JAW_universal_udtb_subsub1.log

\begin{table}[!hb]
{\small
\hspace*{-5mm}
\caption{Comparison of $J^{udtb}_{subsub,1}\label{table3}$}
\begin{tabular}{||c|c|c||}
\hline
\hline
 $\cos\theta$  & $\sqrt{s}=200$GeV                & $\sqrt{s}=10^4$GeV                   \\
\hline
-0.999 &-2.07194259988,   ~~~~7.01180433443       &-0.102517409116E-2,  0.102816191182E-2\\
%                    |12th                |12th                 |12th               |12th
       &-2.07194259988,   ~~~~7.01180433443       &-0.102517409116E-2,  0.102816191182E-2\\
\hline
0      & 0.878509046507E-3,   0.238029806504E-02  & 0.895389809985E-6,  0.184388800385E-6\\
%                     |12th                |12th                 |12th               |12th
       & 0.878509046507E-3,   0.238029806504E-02  & 0.895389809985E-6,  0.184388800385E-6\\
\hline
0.999  & 0.602158734688E-3,   0.913885117129E-03  & 0.517213769582E-6,  0.486927343347E-7\\
%                    |11th                 |12th                 |12th               |12th
       & 0.602158734687E-3,   0.913885117129E-03  & 0.517213769582E-6,  0.486927343347E-7\\
\hline
\hline
\end{tabular}
}
\end{table}
%\vspace*{-2mm}

\noindent
As is seen from this Table, there is agreement within 11-12 digits
for the real parts and within 12 digits for imaginary
parts. We note again that Eq.~(\ref{Jsubsub1_udtb_answ}) derived in this paper
is very compact and  does not contain the masses $\mgm,~\muno,~\mqat$ explicitly.

\subsubsection{No second subtraction $J^{udtb}_{subsub,2}$}
%--------------------------------------------------
%     m1=mtp, m2=mbt, m3=mw, m4=mq-singular
The function $J^{udtb}_{subsub,2}(\Qs,\Ps;\mtp,\mbt,\mw)$ with exchanged arguments has no $\mbt$ mass
singula\-rity. Since we need no limit in $\mtp$, we just rename the second function,
assuming that $\mbt$ is set to zero:
\bqa
J^{udtb}_{subsub,2}(\Qs,\Ps;\mtp,\mbt,\mw)=J^{udtb}_{sub}(\Qs,\Ps;\mtp,\mbt,\mw)\,.
\label{Jsubsub2_udtb_def}
\eqa
The answer is found straightforwardly from Eq.~(\ref{Jsubudtb}) for $J^{udtb}_{sub}$ and $J_0$ from
Eq.~(\ref{J0udtb}) with interchanged arguments $m_1=\mtp$,  $m_2=\mbt=0$:
\bqa
J^{udtb}_{subsub,2} &=& \frac{1}{\Qs+m_1^2}\biggr[
           +\Litwo\left(\frac{\Qs}{m_1^2}\frac{\Delta_{13}}{\P}\right)
           -\Litwo\left(-\frac{\Delta_{13}}{\P}\right)
           -\Litwo\left(-\frac{\Qs}{m_1^2-\ieps}\right)
           +\zeta(2)           \biggl]
\nll &&
              +\frac{1}{\Ps+m_1^2} \biggl[
           -\ln\left(-\frac{\P}{\Ps+\Delta_{13}}\right)
            \ln\left(\frac{m_1^2}{\Ps+m_1^2}\right)
           -\ln\left( R_3 \right) \ln\left(-\frac{\Qs}{\Ps+m_1^2}\right)
\nll &&
           +\Litwo\left(-\frac{\Delta_{13}}{m_3^2-\ieps}\right)
           -\Litwo\left( R_3 \right)
           -\Litwo\left(\frac{\Ps+m_1^2}{m_1^2-\ieps} \right)
\nll &&
           +\Litwo\left(\frac{\Ps+m_1^2}{m_1^2} \frac{1}{1+r_{13}} \right)
           -\Litwo\left(\frac{1}     {1+r_{13}} \right)
           +\zeta(2)      \biggr].
\label{Jsubsub2_udtb_answ}
\eqa
The meaning of $m_3=\mw$ remains unchanged.

In Table~\ref{table4} we give a similar comparison as in Table~\ref{table3}
 but now for the function $J^{udtb}_{subsub,2}$
defined by Eq.~(\ref{Jsubsub2_udtb_def}) and the related ones.
 The setup is the same as for Table~\ref{table1}.
Again, first lines  --- LoopTools, second lines --- this paper.
%\vspace*{-3mm}

%/Export_JAW_udtb_05_12_2008_r8_paper
%Make_JAW_universal_udtb_subsub_2
%JAW_universal_udtb_subsub_2.F
%JAW_universal_udtb_subsub2.log

\begin{table}[!hb]
{\small
\hspace*{-5mm}
\caption{Comparison of $J^{udtb}_{subsub,2}\label{table4}$}
\begin{tabular}{||c|c|c||}
\hline
\hline
 $\cos\theta$  & $\sqrt{s}=200$GeV               & $\sqrt{s}=10^4$GeV                   \\
\hline
-0.999 & 0.159231105282E-4,   0.118244938272E-3  &-0.195869068435E-3  0.240805881821E-3 \\
%                     |12th                |12th                |12th              |12th
       & 0.159231105282E-4,   0.118244938272E-3  &-0.195869068435E-3  0.240805881821E-3 \\
\hline
0      & 0.207059680614E-4,   0.881532449471E-4  & 0.110155292282E-5 -0.167170803336E-6 \\
%                     |12th                 |12th               |12th              |12th
       & 0.207059680614E-4,   0.881532449471E-4  & 0.110155292282E-5 -0.167170803336E-6 \\
\hline
0.999  & 0.234082370485E-4,   0.682484139707E-4  & 0.459502689190E-6 -0.184916952415E-10\\
%                     |12th                |12th                |12th             |11th
       & 0.234082370486E-4,   0.682484139707E-4  & 0.459502689190E-6 -0.184916952411E-10\\
\hline
\hline
\end{tabular}
}
\end{table}
%\vspace*{-2mm}

\noindent
As is seen from this Table, there is again agreement within 11-12 digits for real and imaginary parts.

\section{Functions $J$ for the process $bu \rightarrow td$\label{t-channel}}
%---------------------------------------------------------------------------
In this section we briefly consider $J$ functions arising in four $bu \rightarrow td$ processes.
If one neglects the $b$ quark mass, there appear only four doubly subtracted $J^{butd}_{subsub}$ functions,
see~\cite{Bardin:2009wv}:
\bqa
J^{butd}_{subsub,1}(\Qs,\Ts,0,\mtp,\mw),
\nll
J^{butd}_{subsub,1}(\Qs,\Us,0,\mtp,\mw),
\nll
J^{butd}_{subsub,2}(\Qs,\Ts,\mtp,0,\mw),
\nll
J^{butd}_{subsub,2}(\Qs,\Us,\mtp,0,\mw).
\label{JAW-t-channel}
\eqa

For four possible channels, one has the following correspondence between arguments $\Qs,\Ts,\Us$ and $s,\cos\theta$,
where always $\theta = \angle(\vec{p}_2,\vec{p}_3)$ (see 4-momenta assignment below):
        
\begin{enumerate}       
\item $b(p_1)+u(p_2) \to d(p_3)+t(p_4)$         
\bq
 \Qs = \frac{s-\mtp^2}{2}\cm\,=\,-t\,,\qquad
 \Ts = \frac{s-\mtp^2}{2}\cp\,=\,-u\,,\qquad
 \Us = -s\,,
\eq
\item $b(p_1)+\bar{d}(p_2)\to\bar{u}(p_3)+t(p_4)$
\bq
 \Qs = \frac{s-\mtp^2}{2}\cm\,=\,-t\,,\qquad
 \Ts = -s\,,\qquad
 \Us = \frac{s-\mtp^2}{2}\cp\,=\,-u\,,
\eq
\item $\bar{b}(p_1)+\bar{u}(p_2) \to \bar{d}(p_3)+\bar{t}(p_4)$
\bq
 \Qs = \frac{s-\mtp^2}{2}\cm\,=\,-t\,,\qquad
 \Ts = \frac{s-\mtp^2}{2}\cp\,=\,-u\,,\qquad
 \Us = -s\,,
\eq
\item $\bar{b}(p_1)+d(p_2)\to u(p_3)+\bar{t}(p_4)$
\bq                             
 \Qs = \frac{s-\mtp^2}{2}\cm\,=\,-t\,,\qquad
 \Ts = -s\,\qquad
 \Us = \frac{s-\mtp^2}{2}\cp\,=\,-u\,,
\label{correspondence}
\eq
\end{enumerate}
where
\bq
c_{\pm}=1\pm\cos\theta.
\eq

There is an important difference between $J$ functions considered in Sections~\ref{decay}
and~\ref{s-channel} and in this Section. 
In the first two cases two Mandelstam variables $\Ts=-t$ and
$\Us=-u$ have the same sign; they are different only by the sign of $\cos\theta$. 
In this, third case, $\Qs$ has always the sense of $-t$, while $\Ts$ and
$\Us$ change their meaning $-u$ or $-s$ and therefore the sign.
This is the reason why two $t$ channel $J$'s, for which both arguments are positive, can not be
computed using the $J$ functions derived for channels considered previously,
and the calculation for such sign assignments has to be repeated from scratch.
% conservation low: mtp^2+Qs+Ts+Us=0
% decay             mtp^2=s+t+u; s,t,u - positive; Qs,Ts,Us - negative;
% s-channel         s=mtp^2-t-u; s-positive, t,u - negative; Qs - negative; Ts,Us - positive;
% t-channel         Qs=-t - positive; Ts,Us - channel dependent signs;
This task is beyond the scope of this paper and will be presented elsewhere. 
For the time being we will use a pragmatic solution. 
We noted that equations of Section~\ref{decay} give correct answers for the case $\Qs > 0$ and
$\Ts < 0$ or $\Us < 0$.

%/Export_JAW_butd_28_06_2009_r8_paper
%Make_JAWsubsub_mb1
%main_JAWsubsub_mb1.F
%     JAWsubsub_mb1.f
%mb1_Ps_neg.log
%mb1_Ps_pos.log
%
%Make_JAWsubsub_mb2
%main_JAWsubsub_mb2.F
%     JAWsubsub_mb2.f
%mb2_Ps_neg.log
%mb2_Ps_pos.log

\begin{table}[!hb]
{\small
\hspace*{-5mm}
\caption{Comparison of $J^{butd,-}_{subsub,1}$\label{table5}}
\begin{tabular}{||c|c|c||}
\hline
\hline
 $\cos\theta$  & $\sqrt{s}=200$GeV                & $\sqrt{s}=10^4$GeV                       \\
\hline
-0.999 &-0.173090211233E-04,   0.216758293310E-04 & -0.467497992886E-06,   0.254446155152E-06\\
%                    |11                   |11                  |10                   |10
       &-0.173090211232E-04,   0.216758293312E-04 & -0.467497992836E-06,   0.254446155157E-06\\
\hline
0      &-0.451130695567E-05,   0.149343040736E-05 & -0.462653095799E-06,   0.210944911427E-06\\
%                    |11                  |10                   |10                    |11
       &-0.451130695569E-05,   0.149343040746E-05 & -0.462653095748E-06,   0.210944911431E-06\\
\hline
0.999  & 0.172894787644E-04,  -0.216698015838E-04 &  0.105671318711E-06,  -0.193311004466E-06\\
%                    |11                   |11                  |10                    |11
       & 0.172894787643E-04,  -0.216698015840E-04 &  0.105671318762E-06,  -0.193311004461E-06\\
\hline
\hline
\end{tabular}
}
\end{table}

This is illustrated by Tables~\ref{table5}--\ref{table6}, where
we give a comparison of real and imaginary parts of the function $J^{butd,-}_{subsub,1}$
 and $J^{butd,-}_{subsub,2}$ in the $t$ channel,
Eq.~(\ref{JAW-t-channel}), for the case $\Qs > 0$ and $\Ts < 0$ or $\Us < 0$ using the same formulae and setup as for
Tables~\ref{table1} and~\ref{table2}; first lines  --- LoopTools, second lines --- this paper.

%\clearpage

\begin{table}[!h]
{\small
\hspace*{-5mm}
\caption{Comparison of $J^{butd,-}_{subsub,2}$\label{table6}}
\begin{tabular}{||c|c|c||}
\hline
\hline
 $\cos\theta$  & $\sqrt{s}=200$GeV                & $\sqrt{s}=10^4$GeV                      \\
\hline
-0.999 & 0.108310963092E-03,   0.297113203560E-03 &-0.781547133959E-06,   0.303109233776E-06\\
%                     |12                   |12                 |11                    |12
       & 0.108310963092E-03,   0.297113203560E-03 &-0.781547133958E-06,   0.303109233776E-06\\
\hline
0      & 0.288401492494E-04,   0.656706410374E-04 &-0.723035031957E-06,   0.259579787125E-06\\
%                    |11                   |11                   |12                   |12
       & 0.288401492495E-04,   0.656706410373E-04 &-0.723035031957E-06,   0.259579787125E-06\\
\hline
0.999  &-0.803387429231E-04,  -0.296822348620E-03 & 0.100393594508E-06,  -0.166942822727E-06\\
%                    |11                    |12                  |12                   |12
       &-0.803387429230E-04,  -0.296822348620E-03 & 0.100393594508E-06,  -0.166942822727E-06\\
\hline
\hline
\end{tabular}
}
\end{table}
%\vspace*{-2mm}

As is seen from these Tables, there is agreement within 10-11 digits for real and imaginary parts.
Note that one can reach agreement to all visible digits with Real*16.

As far as two $t$ channel $J$'s are concerned, for which both arguments are positive,
 we accept a temporary solution for the time being, noticing that previously
  computed $J$ functions return correctly only the real parts. We recall that imaginary parts
do not contribute at the one-loop level. The real parts are illustrated by Table~\ref{table7}.

\begin{table}[!h]
\centering
{\small
\caption{Comparison of $J^{butd,+}_{subsub,1(2)}$\label{table7}}
\begin{tabular}{||c|c|c||}
\hline
\hline
 $\cos\theta$  & $\sqrt{s}=10^4$GeV  & $\sqrt{s}=10^4$GeV \\
\hline
               & $J^{butd,+}_{subsub,1}$ &  $J^{butd,+}_{subsub,2}$ \\
\hline
-0.999999 & -0.372108952046D+01 &  -0.122105779653D-02 \\
%                       |10                      |12
          & -0.372108952003D+01 &  -0.122105779653D-02 \\
\hline
-0.9990   & -0.122256223637D-02 &  -0.256945712509D-03 \\
%                         |12                    |12
          & -0.122256223637D-02 &  -0.256945712509D-03 \\
\hline
0         &  0.707331687150D-06 &   0.122739516453D-05 \\
%                       |10                      |12
          &  0.707331687129D-06 &   0.122739516453D-05 \\
\hline
0.9999    & -0.155145904970D-06 &  -0.149761802457D-06 \\
%                       |10                      |12
          & -0.155145905078D-06 &  -0.149761802457D-06 \\
\hline
0.99999   & -0.515690518852D-06 &  -0.496884568518D-06 \\
%                      |9                       |11
          & -0.515690519134D-06 &  -0.496884568519D-06 \\
\hline
\hline
\end{tabular}
}
\end{table}
%\vspace*{-2mm}
As is seen from these Tables, there is an agreement within 9-12 digits for real parts
and that the agreement does not improve with Real*16 computations.

\section{SANC packages}
%----------------------
\begin{figure}[htp!]
\includegraphics[width=\textwidth]{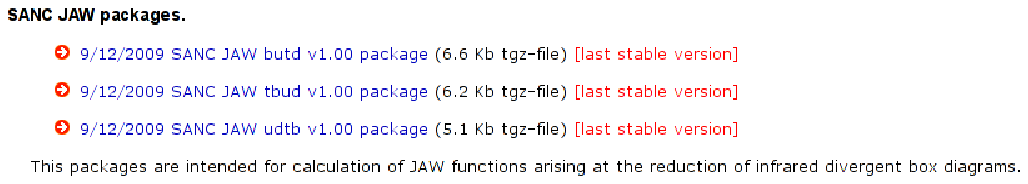}
\caption[{\tt SANC} packages download page]
        {{\tt SANC} packages download page}
\end{figure}

The numeric comparison with the LoopTools library presented 
in this paper can be verified with help of the
SANC software packages. We have three packages related to the $J_{subsub,1(2)}$
functions for three channels. They are available to download from the web
pages of SANC project~\cite{homepagesSANC}.

Each package contains the following files:
\begin{itemize}
\item main files for both $J_{subsub,1(2)}$ functions;
\item $J_{subsub,1(2)}$ source files;
\item utility functions;
\item Makefile;
\item README, INSTALL, LICENSE and other information files.
\end{itemize}

%%% Local Variables: 
%%% mode: latex
%%% TeX-master: t
%%% End: 

\section{Conclusions}
%--------------------

The invention and usage of a
new class of functions $J$, relevant to the Passarino--Veltman reduction~\cite{Passarino:1978jh},
has become a standard step in the chain of calculations in project SANC.
Originally they were introduced in~\cite{Bardin:2009zz}.

 Obvious advantages and disadvantages of these functions are:
\begin{enumerate}
\item final results are compact and evidently demonstrate the fundamentals of physics;
\item the subtracted $J_{WA,\rm{sub}}$ functions have no mass singularities;
\item their compactness results in stable and very rapid calculations;
\item the  functions of this new class are linear combinations of the standard Passarino-Veltman
$D_0$ and $C_0$ functions and their explicit form depends on the concrete channel of a process;
namely, they depend of the choice of channel
 because we have no universal expressions for the $J$ functions,
\item for the analytic calculations we have, nevertheless, a consistent method;
\item it would be desirable to improve the way of the analytic calculations in order to obtain various channels
by a simple rotation of their arguments, as is done in the LoopTools package for $D_0$ functions.
\end{enumerate}

\section*{Acknowledgement}
We are indebted to S.G.~Bondarenko for useful discussions.
This paper is partly supported by grant RFFI $N^{o}$ 07-02-00932.

\bibliographystyle{utphys_spires}
\addcontentsline{toc}{section}{\refname}\bibliography{bib}

\providecommand{\href}[2]{#2}\begingroup\begin{thebibliography}{1}

\bibitem{Passarino:1978jh}
G.~Passarino and M.~J.~G. Veltman, {\em Nucl. Phys.} {\bf B160} (1979)
151.
%%CITATION = NUPHA,B160,151;%%.

\bibitem{Bardin:1999ak}
D.~Y. Bardin and G.~Passarino, Oxford, UK: Clarendon (1999) 685 p.

\bibitem{Bardin:2009zz}
D.~Y. Bardin, L.~V. Kalinovskaya, and L.~A. Rumyantsev, {\em Phys. Part. Nucl.
  Lett.} {\bf 6} (2009)
30--41.
%%CITATION = 00438,6,30;%%.

\bibitem{Bardin:2009wv}
D.~Bardin {\em et al.}, {\em Phys. Part. Nucl. Lett.} {\bf 7} (2010) 128--141,
\href{http://www.arXiv.org/abs/0903.1533 [hep-ph]}{{\tt 0903.1533 [hep-ph]}}.
%%CITATION = 0903.1533;%%.

\bibitem{Hahn:1998yk}
T.~Hahn and M.~Perez-Victoria, {\em Comput. Phys. Commun.} {\bf 118} (1999)
  153--165,
\href{http://www.arXiv.org/abs/hep-ph/9807565}{{\tt hep-ph/9807565}}.
%%CITATION = HEP-PH/9807565;%%.

\bibitem{homepagesSANC}
{\em Dubna --- {\em http://sanc.jinr.ru}, CERN --- {\em
  http://pcphsanc.cern.ch}} (2007).

\end{thebibliography}\endgroup

\end{document}